\pdfoutput=1

\documentclass[11pt]{article}

\usepackage[final]{acl}

\usepackage{times}
\usepackage{latexsym}
\usepackage{hyperref}

\usepackage[T1]{fontenc}

\usepackage[utf8]{inputenc}

\usepackage{microtype}

\usepackage{inconsolata}

\usepackage{graphicx}
\usepackage{multirow}
\usepackage{fontawesome5}

\usepackage{subcaption}
\usepackage{adjustbox}
\usepackage[most]{tcolorbox}
\newtcolorbox[auto counter,number within=section]{pabox}[2][]{%
colback=white!5!white,colframe=blue!60!green,fonttitle=\small\bfseries,
title=Prompt.~\thetcbcounter: #2,#1}
\usepackage{booktabs}
\usepackage{colortbl}

\usepackage{array}           
\usepackage{tabularx}       
\usepackage{adjustbox} 
\usepackage{wrapfig}

\title{SERVAL: Surprisingly Effective Zero-Shot Visual Document Retrieval \\ Powered  by Large Vision and Language Models}

\author{
  Thong Nguyen$^{1}$ \quad
  Yibin Lei$^{1}$ \quad
  Jia-Huei Ju$^{1}$ \quad
  Andrew Yates$^{2}$ \\
  $^{1}$University of Amsterdam \\
  $^{2}$Johns Hopkins University \\
  \texttt{\{t.nguyen2,y.lei,j.ju\}@uva.nl}, \texttt{andrew.yates@jhu.edu}
}

\begin{document}
\maketitle
\begin{abstract}

Visual Document Retrieval (VDR) typically operates as text-to-image retrieval using specialized bi-encoders trained to directly embed document images. We revisit a zero-shot generate-and-encode pipeline: a vision–language model first produces a detailed textual description of each document image, which is then embedded by a standard text encoder. On the ViDoRe-v2 benchmark, the method reaches 63.4\% nDCG@5, surpassing the strongest specialised multi-vector visual document encoder. It also scales better to large collections and offers broader multilingual coverage. Analysis shows that modern vision–language models capture complex textual and visual cues with sufficient granularity to act as a reusable semantic proxy. By offloading modality alignment to pretrained vision–language models, our approach removes the need for computationally intensive text-image contrastive training and establishes a strong zero-shot baseline for future VDR systems. Our code is available for reproduction at: \href{https://github.com/thongnt99/serval}{\faGithub\ thongnt99/serval}

\end{abstract}

\section{Introduction and Related Work}

Real‑world documents originate from diverse sources, spanning the public web to private enterprise repositories, and appear in many formats, including plain text, figures, graphs, and tables. Document retrieval bridges human or artificial agents to the most relevant information, enabling informed decision‑making and knowledge synthesis.
\begin{figure}[ht!]
    \centering
    \includegraphics[width=1.0\linewidth]{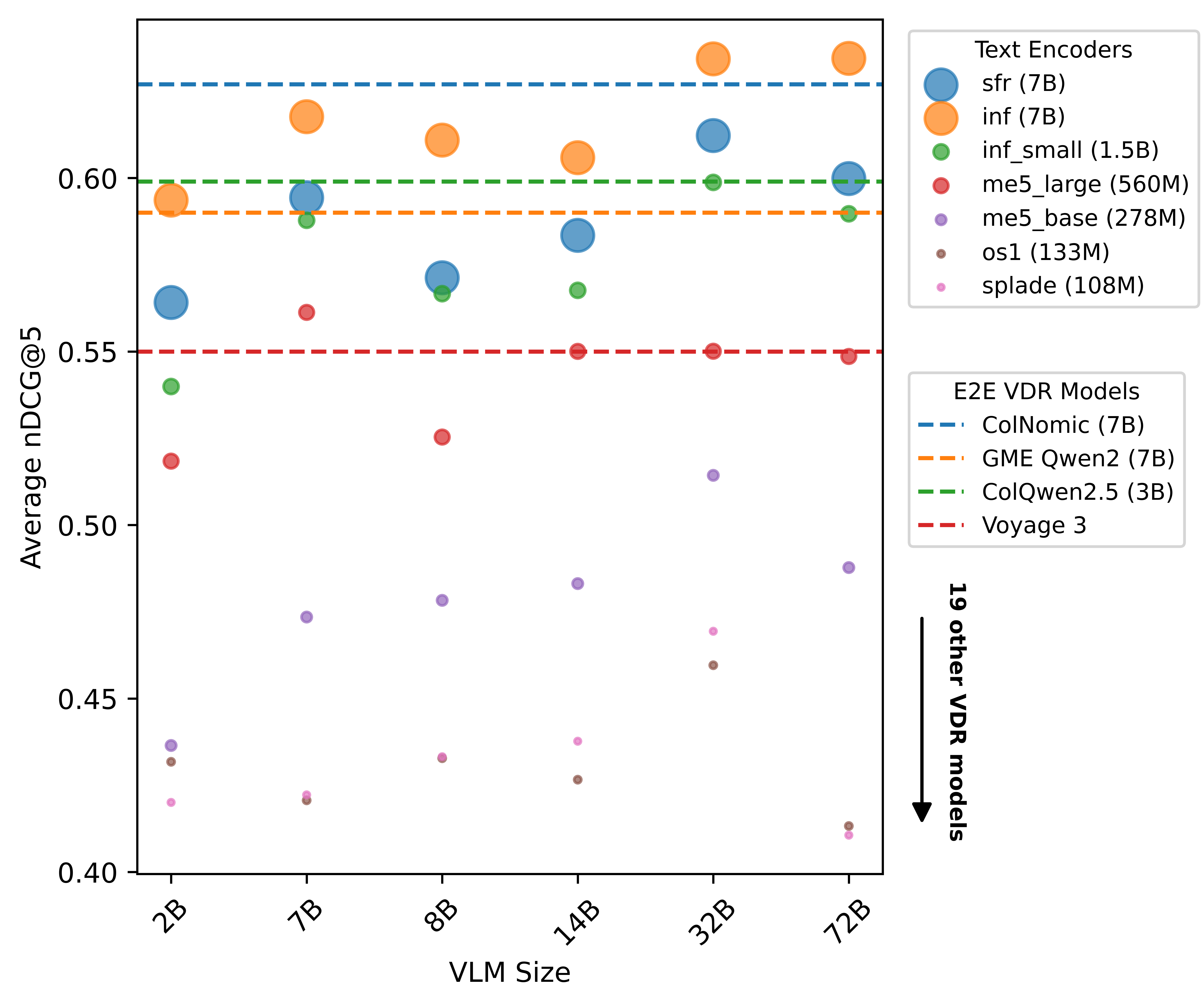}
    \caption{nDCG@5 for zero-shot Visual Document Retrieval using VLMs and text encoders of varying scales. Despite no task-specific training, our zero-shot method could compete with end-to-end models explicitly trained for VDR on large-scale text-(document) image datasets.}
    \label{fig:compare_ndcg@5}
    \vspace{-3mm}
\end{figure}

Over several decades the field has been dominated by text‑centric retrieval methods. Classic approaches such as BM25~\cite{robertson1994some} rely purely on lexical matching between queries and documents. The recent advent of deep learning, and especially transformer architectures such as BERT, has shifted the focus to context‑aware neural retrieval. Today a rich ecosystem of neural paradigms is available, from sparse and dense first‑stage encoders~\cite{wang2022text,wang2024multilingual,formal2021splade,lassance2024splade,nguyen2023unified} to cross‑encoders for re‑ranking~\cite{nogueira2020document,macavaney2019cedr,reimers-2019-sentence-bert}. Trained on large human‑ and machine‑annotated corpora, these models achieve state‑of‑the‑art performance on in‑domain~\cite{bajaj2016ms}, out‑of‑domain~\cite{thakur2021beir}, and multilingual~\cite{zhang2023miracl,enevoldsen2025mmteb,zhang2021mr} benchmarks.

Text‑only retrieval overlooks other important visual elements embedded in documents~\cite{xu2020layoutlm}. Visual Document Retrieval (VDR) tackles this gap by jointly encoding textual and visual elements rendered as images. \citet{ma2024unifying} formulate VDR as text‑to‑image retrieval by rasterizing each page, while CoPali introduces a multi‑vector encoder together with the ViDoRe benchmark~\cite{faysse2024colpali}. To spur further progress, ViDoRe‑v2~\cite{mace2025vidore} raises the bar with human‑verified, multilingual queries that are substantially more challenging.

To close the modality gap, state-of-the-art VDR systems typically rely on expensive contrastive training over high-quality text and visual document pairs. However, we hypothesize that advanced vision-language models (VLMs) can bridge this gap by effectively describing visual elements in language. Therefore, we reformulate the end-to-end paradigm and revisit a simple zero-shot alternative that decouples the problem into two independent sub-tasks: (i) document-description generation, handled by a VLM, and (ii) text encoding, handled by a conventional pretrained text encoder. This modular design enables us to plug in best-in-class components and leverage the growing number of high-quality VLMs and text encoders.

For description generation, we exploit recent VLMs, including Qwen2.5VL~\cite{bai2025qwen2} and InternVL 3~\cite{zhu2025internvl3}, that excel at visual understanding. For text encoding we leverage robust open‑source multilingual encoders~\cite{wang2022text,wang2024multilingual,SFRAIResearch2024, infly-ai_2025} usually equipped with instruction‑following capabilities~\cite{weller2024promptriever}.

Evaluated on the nine ViDoRe‑v2 tasks and MIRACL-VISION benchmark, our zero-shot generate-and-encode approach matches or even surpasses state-of-the-art supervised multi‑vector baselines (as shown in Figure~\ref{fig:compare_ndcg@5}), despite using no VDR‑specific training data. Our analysis produces three key insights:
\begin{itemize}
\item Recent vision–language models (VLMs) can accurately describe visual elements embedded in documents--figures, graphs, and tables--enabling effective visual document retrieval from their generated descriptions.
\item Scaling both VLMs and text encoders improves retrieval performance, but scaling/improving the text encoder yields better gains. Large (32B / 72B) VLMs paired with strong text encoders achieve the best scores, while even 2B–7B models already surpass most supervised end-to-end models on ViDoRe-v2.
\item Supervised end-to-end VDR models perform well on English, but they lag behind on multilingual and cross-lingual tasks, highlighting potential room for future improvement.
\end{itemize}

Because description generation is executed offline during indexing, online latency remains similar to end-to-end VDR approaches. Smaller VLMs (2B–7B) offer an attractive speed–accuracy trade‑off, providing competitive accuracy while reducing offline document preprocessing cost.




\section{Zero-shot Visual Document Retrieval}
\begin{figure}[ht!]
    \centering
    \includegraphics[width=1.0\linewidth]{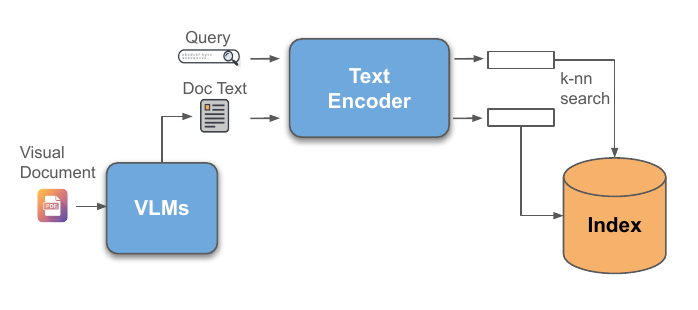}
    \caption{Zero-shot VDR using VLM-generated document descriptions and a pretrained text encoder.}
    \label{fig:zeroshot-vdr-pipeline}
\end{figure}
\begin{figure*}[t]
    \centering
    \includegraphics[width=0.8\linewidth]{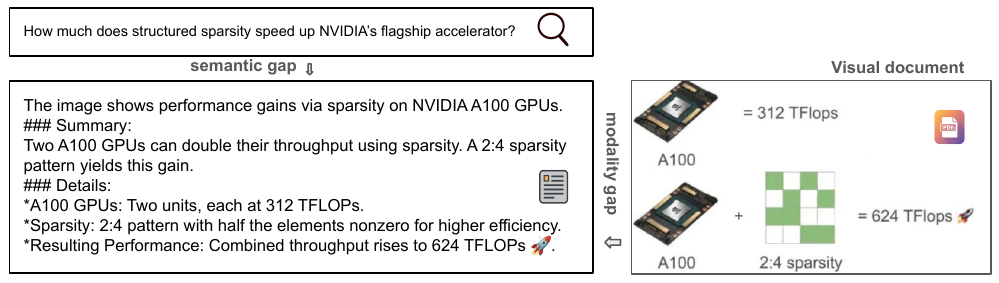}
    \caption{Example of a document description generated by InternVL3. Beyond simple OCR, the model integrates textual and visual cues into natural language, bridging the modality gap.}
    \label{fig:demo_doc_description}
    \vspace{-1mm}
\end{figure*}

The workflow for our zero-shot VDR approach is illustrated in Figure \ref{fig:zeroshot-vdr-pipeline}. First, a vision-language model (VLM) is used to generate a detailed description of the visual document. Using the prompt shown in Prompt \ref{vlm_prompt}, we instruct the VLM to begin with an overall summary of the content depicted in the image, followed by a comprehensive list of details, including any extracted text and numerical values. The entire description is generated in a single step, bypassing intermediate procedures such as layout detection or document chunking used in CoPali~\cite{faysse2024colpali}. An example of a generated description from InternVL3 is shown in Figure \ref{fig:demo_doc_description}. Additional examples covering various complex document types (e.g., graphs, tables, diagrams) are provided in the Appendix \ref{sec:generated_descriptions}.

\begin{pabox}[label={vlm_prompt}, width=\columnwidth]{Description generation with VLMs}
Provide a comprehensive description of the document in the image in English. Begin with a summary, then follow with details. Extract all visible text and numerical values from the document.
\end{pabox}

In the second step, any off-the-shelf text encoder can be used to encode both queries and the generated document descriptions. The goal of this encoder is to bridge the semantic gap between queries and documents by mapping them into a shared semantic space where relevant pairs are positioned closely. It is also important for the encoder to support multilingual and cross-lingual retrieval, as the evaluation will include non-English data. For large-scale document collections, document descriptions and their embeddings can be pre-computed and indexed offline in a database.

\section{Experimental Setup}
We evaluate our zero-shot VDR approach and baselines on the ViDoRe-v2 benchmark~\cite{mace2025vidore}, an enhanced and more challenging successor to ViDoRe-v1. ViDoRe-v2 comprises nine VDR tasks spanning four languages (English, Spanish, French, and German) and covers a variety of domains, including business, restaurant, and medical. It features real-world, complex queries, for which the best end-to-end supervised VDR model~\cite{nomicembedmultimodal2025} only achieves a moderate nDCG@5 of 0.62. The document collection includes both text-centric items (e.g., reports composed primarily of text) and visually rich items (e.g., slide decks containing tables or graphs).

While Vidore-V2 provides a strong basis for evaluating multilingual multimodal retrieval, its collections remain relatively limited in scale and language coverage. To complement these experiments, we further evaluate on MIRACL-VISION~\cite{osmulski2025miracl}, a benchmark that extends MIRACL with vision-language data and offers larger multilingual collections as well as broader coverage of both high- and low-resource languages. This makes MIRACL-VISION a more realistic testbed for assessing the generalization ability of retrieval models beyond the settings captured in Vidore-V2.

To generate document descriptions, we use various VLMs, including Qwen2.5VL (7B, 32B, and 72B)~\cite{bai2025qwen2} and InternVL3 (2B and 8B)~\cite{zhu2025internvl3}. For all models, we use quantized versions with Activation-Aware Weight Quantization and accelerate inference using the vLLM~\cite{kwon2023efficient} and LMDeploy~\cite{2023lmdeploy} frameworks. We provide statistics on the token length of the generated descriptions in Table \ref{tab:vlm_tokens}. The average number of tokens generated by different VLMs is roughly around 500 tokens per document, with the exception of QwenVL2.5 32B that produces about 1000 tokens/doc.

\begin{table}[ht!]
\centering
\begin{tabular}{l r}
\toprule
\toprule
\textbf{Models} & \textbf{\# tokens per doc} \\
\midrule
QwenVL2.5 -- 7B   & 422.62 \\
QwenVL2.5 -- 32B  & 1009.52 \\
QwenVL2.5 -- 72B  & 636.03 \\
InternVL3 -- 2B   & 515.53 \\
InternVL3 -- 8B   & 619.13 \\
InternVL3 -- 14B  & 537.11 \\
\bottomrule
\bottomrule
\end{tabular}
\caption{Average number of tokens generated when producing visual document descriptions.}
\label{tab:vlm_tokens}
\end{table}

For text encoders, we experiment with two families: learned sparse retrieval and dense retrieval. For sparse retrieval, we employ Splade-v3~\cite{lassance2024splade} and the Open Search sparse model~\cite{geng2024towards}. For dense retrieval, we evaluate a range of multilingual and instruction-tuned models, including: \textit{multilingual-e5-base}, \textit{multilingual-e5-large}, \textit{SFR-Embedding-Mistral}, and \textit{inf-retriever-v1}. All sparse and dense retrieval checkpoints are publicly available on HuggingFace~\cite{wolf-etal-2020-transformers}.

We report nDCG@k and Recall@k (R@k) for $k = \{1, 5, 10\}$, consistent with the evaluation protocol used in prior ViDoRe benchmarks. 

\section{Results and Discussion}

\begin{table*}[ht!]
  \centering
  \tiny 
  \setlength{\tabcolsep}{3pt}
  \begin{tabular}{l l c c c c c c c c c | c}
\toprule
\toprule
\textbf{VLM} & \textbf{Text Encoder} & \textbf{RERB} & \textbf{SAXA} & \textbf{SAXAM} & \textbf{SEME} & \textbf{SMBTI} & \textbf{SMBTIM} & \textbf{SRS} & \textbf{SRSM} & \textbf{SEMEM} & \textbf{AVG} \\
\midrule
\addlinespace[1ex]
\rowcolor{gray!10}
\rowcolor{gray!10} \multicolumn{2}{l}{ColNomic Embed Multimodal 7B~\cite{nomicembedmultimodal2025}} & \textbf{73.9} & 68.3 & \textbf{61.3} & 61.6 & \textbf{66.1} & \textbf{64.2} & 54.7 & \textbf{57.3} & 56.7 & \textbf{62.7} \\
\rowcolor{gray!10} \multicolumn{2}{l}{ColNomic Embed Multimodal 3B~\cite{nomicembedmultimodal2025}} & 65.8 & 68.8 & 61.0 & 60.2 & 63.5 & 62.5 & 55.4 & 56.6 & \textbf{57.2} & 61.2 \\
\rowcolor{gray!10} \multicolumn{2}{l}{T-Systems ColQwen2.5-3B~\cite{faysse2024colpali}} & 72.1 & \textbf{69.3} & 60.0 & 54.8 & 65.3 & 61.7 & 51.2 & 51.7 & 53.3 & 59.9 \\ 
\rowcolor{gray!10} \multicolumn{2}{l}{GME Qwen2 7B~\cite{zhang2024gme}}                      & 65.8 & 60.7 & 55.4 & \textbf{62.9} & 64.0 & 55.1 & \textbf{56.2} & 54.3 & 56.7 & 59.0 \\

\rowcolor{gray!10} \multicolumn{2}{l}{Voyage Multimodal 3~\cite{voyage-multimodal-3}} & 56.1 & 64.1 & 59.5 & 58.8 & 56.4 & 51.5 & 55.0 & 47.2 & 46.2 & 55.0 \\ 
\midrule
\multirow{4}{*}{Qwen2.5VL-7B} & splade 108M~\cite{lassance2024splade} & 64.8 & 32.1 & 30.6 & 56.4 & 59.8 & 39.8 & 33.8 & 35.4 & 27.4 & 42.2 \\
 & me5\_large 560M~\cite{wang2024multilingual} & 64.5 & 56.5 & 58.0 & 52.5 & 58.6 & 55.9 & 55.9 & 55.5 & 47.8 & 56.1 \\
 & inf\_small 1.5B~\cite{infly-ai_2025} & 60.2 & 65.2 & 64.4 & \textbf{58.1} & 60.6 & 57.1 & 57.1 & 54.9 & 51.5 & 58.8 \\
 & inf 7B~\cite{infly-ai_2025} & \textbf{66.3} & \textbf{72.0} & \textbf{65.0} & 57.2 & \textbf{64.9} & \textbf{62.2} & \textbf{57.4} & \textbf{56.3} & \textbf{54.5} & \textbf{61.8} \\
\midrule
\multirow{4}{*}{Qwen2.5VL-32B} & splade 108M~\cite{lassance2024splade} & 69.3 & 53.1 & 38.5 & 58.1 & 63.1 & 40.8 & 35.0 & 35.2 & 29.4 & 46.9 \\
 & me5\_large 560M~\cite{wang2024multilingual} & 68.8 & 53.8 & 53.5 & 49.1 & 60.6 & 57.1 & 52.0 & 54.0 & 46.2 & 55.0 \\
 & inf\_small 1.5B~\cite{infly-ai_2025} & 67.9 & 68.3 & 66.4 & 56.6 & 60.9 & 58.6 & 54.8 & 55.3 & 50.0 & 59.9 \\
 & inf 7B~\cite{infly-ai_2025} & \textbf{70.7} & \textbf{69.7} & \textbf{69.4} & \textbf{59.0} & \textbf{65.1} &\textbf{ 63.0} & \textbf{59.0} & \textbf{59.5} & \textbf{55.5} & \textbf{63.4} \\
\midrule
\multirow{4}{*}{Qwen2.5VL-72B} & splade 108B~\cite{lassance2024splade} & 63.8 & 16.4 & 25.8 & 59.8 & 62.7 & 42.1 & 31.9 & 37.0 & 30.1 & 41.1 \\
 & me5\_large 560M~\cite{wang2024multilingual} & 66.1 & 50.8 & 56.0 & 52.5 & 61.6 & 58.0 & 50.5 & 49.0 & 49.2 & 54.9 \\
 & inf\_small 1.5B~\cite{infly-ai_2025} & 67.3 & 58.8 & 62.1 & 62.7 & 62.6 & 59.3 & 51.7 & 53.7 & 52.6 & 59.0 \\
 & inf 7B~\cite{infly-ai_2025} & \textbf{68.9} & \textbf{70.7} & \textbf{67.9} & \textbf{60.3} & \textbf{65.5} & \textbf{63.2} & \textbf{58.2} & \textbf{58.6} & \textbf{57.7} & \textbf{63.4 }\\
\midrule
\multirow{4}{*}{InternVL3-2B} & splade 108M~\cite{lassance2024splade} & 52.7 & 48.1 & 28.6 & 58.6 & 58.6 & 39.6 & 28.4 & 33.5 & 30.0 & 42.0 \\
 & me5\_large 560M~\cite{wang2024multilingual} & 61.4 & 52.9 & 42.0 & 51.4 & 56.7 & 54.8 & 52.6 & 50.7 & 43.9 & 51.8 \\
 & inf\_small 1.5B ~\cite{infly-ai_2025} & 59.8 & 62.5 & 54.1 & 52.9 & 59.3 & 55.2 & 47.2 & 48.7 & 46.2 & 54.0 \\
 & inf 7B~\cite{infly-ai_2025} & \textbf{63.3} & \textbf{72.5} & \textbf{62.6} & \textbf{55.0} & \textbf{63.8} & \textbf{60.7} & \textbf{52.9} & \textbf{52.1} & \textbf{51.3} & \textbf{59.4} \\
\midrule
\multirow{4}{*}{InternVL3-8B} & splade 108M ~\cite{lassance2024splade} & 65.6 & 30.7 & 31.2 & 58.0 & 60.6 & 40.6 & 34.9 & 38.2 & 30.3 & 43.3 \\
 & me5\_large 560M ~\cite{wang2024multilingual} & 58.1 & 38.8 & 49.6 & 52.2 & 59.8 & 57.7 & 53.1 & 56.0 & 47.5 & 52.5 \\
 & inf\_small 1.5B~\cite{infly-ai_2025} & 63.0 & 54.4 & 57.6 & \textbf{58.8} & 63.4 & 57.8 & 50.6 & 51.9 & 52.5 & 56.7 \\
 & inf 7B~\cite{infly-ai_2025} & \textbf{66.7} & \textbf{62.9 }& \textbf{64.8} & 57.3 & \textbf{65.5 }& \textbf{63.0} & \textbf{57.6} & \textbf{57.4 }& \textbf{54.6} & \textbf{61.1} \\
\midrule
\multirow{4}{*}{InternVL3-14B} & splade 108M~\cite{lassance2024splade} & \textbf{68.5} & 38.0 & 31.4 & 53.8 & 61.0 & 40.7 & 36.9 & 37.6 & 26.1 & 43.8 \\
 & me5\_large 560M~\cite{wang2024multilingual} & 67.7 & 54.3 & 53.0 & 52.4 & 58.3 & 55.2 & \textbf{54.2} & \textbf{55.1} & 44.9 & 55.0 \\
 & inf\_small 1.5B~\cite{infly-ai_2025} & 60.5 & 60.1 & 57.3 & 58.0 & 62.4 & 57.6 & 51.2 & 52.1 & 51.7 & 56.8 \\
 & inf 7B~\cite{infly-ai_2025} & 67.7 & \textbf{64.7} & \textbf{63.0} & \textbf{58.9 }& \textbf{64.7} & \textbf{62.7} & 52.4 & \textbf{54.6 }& \textbf{56.5} & \textbf{60.6} \\
\bottomrule
\bottomrule
\end{tabular}
  \caption{Zeroshot VDR, by encoding generated descriptions with single-vector encoders, performs competitively with SOTA multi-vector end-to-end VDR approaches (in \colorbox{gray!10}{gray}) on ViDoRe-v2 benchmark. Metric: nDCG@5.}
  \label{tab:main_results}
  \vspace{-3mm}
\end{table*}

The main results using our zero-shot approaches and end-to-end VDR baselines are presented in Table~\ref{tab:main_results} and visualized in Figure~\ref{fig:compare_ndcg@5}. Due to space limitations, we primarily use nDCG@5 for analysis. Please refer to Appendix~\ref{sec:detailed_zeroshot_results} for the complete results and Appendix \ref{sec:dataset_abbr} for dataset abbreviations.

Overall, the zero-shot VDR approaches are shown to be highly competitive with strong supervised single-vector and multi-vector VDR baselines (highlighted with a gray background in Table~\ref{tab:main_results}) that directly embed visual documents. The state-of-the-art supervised model, ColNomic Embed Multimodal 7B~\cite{nomicembedmultimodal2025}, achieves an average nDCG@5 of 62.7. In contrast, our best zero-shot method, using Qwen2.5VL (32B, 72B)\cite{bai2025qwen2} for description generation and INF 7B~\cite{infly-ai_2025} for text encoding, achieves an average score of 63.4, surpassing all supervised VDR models on the ViDoRe-v2 benchmark. This result highlights the effectiveness of VLMs in generating rich descriptions for visual documents and the capability of text encoders to close the query-document semantic gap. 

The separation of generation and encoding allows us to flexibly plug and play different models at each stage. We explore a range of generation and text encoding models at varying scales. For generation, we experiment with vision-language models (VLMs) from 2B to 72B parameters, including Qwen2.5VL~\cite{bai2025qwen2} and InternVL3~\cite{zhu2025internvl3}. For text encoding, we evaluate both dense and sparse encoders, from the lightweight BERT~\cite{Devlin2019BERTPO} with 108M parameters to large 7B-scale encoders. As shown in Figure~\ref{fig:compare_ndcg@5}, scaling both VLMs and text encoders leads to notable improvements. With descriptions generated by Qwen2.5VL-32B, the multilingual M5 encoder (560M) achieves a moderate nDCG@5 of 55.0, while replacing M5 with INF-1.5B and INF-7B yields gains of 8.9\% and 15\%, respectively. Interestingly, even with the smallest VLM (InternVL3-2B), the INF-7B encoder achieves strong performance (nDCG@5 = 59.4), outperforming most supervised VDR models on ViDoRe-v2. Using larger VLMs (e.g., 7B or 72B) can further enhance retrieval performance by up to 6\%, although we find that scaling the text encoder provides greater benefit than scaling the VLMs.

Examining dataset-level results reveals contrasting trends between English and non-English corpora. E2E models trained on multilingual data (e.g., ColNomic or ColQwen-2.5) excel on the English-only corpora, including RERB (Restaurants), SMBTI (Biomedical), and SEME (Economics). In contrast, zero-shot approaches, except for Splade-v3~\cite{lassance2024splade} trained exclusively on English, perform better on the multilingual and cross-lingual retrieval tasks, most notably on SAXAM (Insurance) and SRS (Restaurant). These findings underscore the need to further strengthen the multilingual capabilities of end-to-end Visual Document Retrieval models.

\begin{table*}[ht]
\centering
\small
\setlength{\tabcolsep}{3pt} 

\begin{tabular}{lccccccccc}
\toprule
\toprule
\textbf{Model} & \textbf{en} & \textbf{de} & \textbf{ja} & \textbf{zh} & \textbf{fr} & \textbf{yo} & \textbf{sw} & \textbf{id} & \textbf{Avg} \\
\midrule
\rowcolor{gray!10} colqwen2-v1.0                & 64.2 & 60.0 & 69.7 & 49.3 & 68.8 & 51.2 & 49.3 & 53.2 & 58.2 \\
\rowcolor{gray!10} vdr-2b-multi-v1              & 67.8 & 62.1 & 65.5 & 59.6 & 71.9 & 45.8 & 45.1 & 52.5 & 58.8 \\
\rowcolor{gray!10} gme-qwen2-vl2b-instruct      & 67.8 & 63.5 & 73.1 & 63.1 & 68.5 & 48.8 & 53.5 & 54.2 & 61.6 \\
\rowcolor{gray!10} dse-qwen2-2b-mrl-v1          & 66.1 & 62.7 & 62.3 & 59.6 & 71.6 & 41.8 & 41.6 & 48.7 & 56.8 \\
\midrule
InternVL-2B + me5-base       & 61.5 & 62.2 & 72.7 & 60.9 & 69.9 & 67.4 & 63.9 & 56.7 & 64.4 \\
InternVL-2B + me5-large      & 65.5 & 67.3 & 77.3 & 67.0 & 72.8 & 74.1 & 69.0 & 58.0 & 68.9 \\
InternVL-2B + inf-small      & \textbf{71.5} & \textbf{71.3} & \textbf{79.8} & \textbf{75.3} & \textbf{81.3} & \textbf{74.1} & \textbf{69.0} & \textbf{62.6} & \textbf{72.1} \\
\bottomrule
\bottomrule
\end{tabular}
\caption{nDCG@10 performance of SERVAL compared with baselines on MIRACL-VISION.~\cite{osmulski2025miracl}. SERVAL significantly outperforms baselines (in gray) that encode document images directly.}.
\label{tab:miracl-vision}
\end{table*}

As the generate-and-encode process is performed offline on the document side, our zero-shot VDR approach does not add to retrieval latency. Retrieval latency therefore matches that of any end-to-end VDR model that has the same size as our text encoder. Document-side inference is slower because of the description-generation phase, especially with large 32B or 72B VLMs. Techniques such as pruning \cite{sun2023simple}, quantization \cite{lin2024awq}, and K-V caching \cite{li2024survey} can accelerate generation and are already implemented in frameworks like vLLM \cite{kwon2023efficient} and LMDeploy \cite{2023lmdeploy}.

Most importantly, a lightweight configuration that pairs a small VLM (InternVL3-2B) with a small text encoder (INF 1.5B) still achieves an nDCG@5 of 54.0. This matches the performance of the supervised Voyage Multimodal 3 \cite{voyage-multimodal-3}, yet requires no task-specific training, highlighting the cost-efficient appeal of our zeroshot generate-and-encode approach.

Table \ref{tab:miracl-vision} presents results on MIRACL-VISION~\cite{osmulski2025miracl}, a benchmark with larger multilingual collections than Vidore-V2 and a broader coverage of languages, including low-resource ones. This setting tests the scalability of SERVAL beyond small collections. SERVAL (InternVL-2B + inf-small) achieves the best average score of 72.1 nDCG@10, compared to 61.6 for the strongest baseline (gme-qwen2-vl2b-instruct), a gain of +10.5 points. The improvements are particularly pronounced on low-resource languages, with +25.3 on Yoruba (74.1 vs. 48.8) and +15.5 on Swahili (69.0 vs. 53.5). These results demonstrate that SERVAL not only scales effectively to larger collections but also delivers robust performance across both high- and low-resource languages.

\section{Generation Latency}
\begin{table}[ht]
\centering
\small
\begin{tabular}{l c}
\toprule
\toprule
\textbf{Models} & \textbf{s/img} \\
\midrule
QwenVL2.5 -- 7B   & 0.290 \\
QwenVL2.5 -- 32B  & 1.080 \\
QwenVL2.5 -- 72B  & 1.610 \\
InternVL3 -- 2B   & 0.260 \\
InternVL3 -- 8B   & 0.350 \\
InternVL3 -- 14B  & 0.530 \\
\bottomrule
\bottomrule
\end{tabular}
\caption{Average description generation latency when producing descriptions of visual documents.}
\label{tab:inference_time}
\end{table}
We provide an analysis of the time taken to generate a description for a single image with different Vision-Language Models in Table \ref{tab:inference_time}. We measured the generation time on a single H100 GPU using \textbf{lmdeploy}~\cite{2023lmdeploy} as the inference engine. Small VLMs ($\leq$14B parameters) take half a second or less to generate one description, while larger models (i.e., QwenVL2.5 32B and 72B) require more than one second per image. This generation step is not latency-sensitive, as it can be performed offline and only once per document. It is also feasible to use lower-end GPUs like A100s, which would only tolerably increase this offline generation time. The encoding latency of our approach on generated text is the same as (or even faster than, as we skip the vision transformer encoder) that of an end-to-end supervised VDR encoder.

\section{Conclusion}
In this work, we revisit zero-shot visual document retrieval by pairing a pretrained text encoder with document descriptions generated by a vision–language model. Contrary to the findings of \citet{faysse2024colpali}, we show that this simple training-free method achieves strong performance on the recent ViDoRe-v2 and MIRACL-VISION benchmarks, rivaling state-of-the-art end-to-end models trained specifically for VDR. Moreover, our zeroshot generate-and-encode strategy remains robust even in lightweight, practical settings that rely on small VLMs and compact text encoders. We hope this study establishes a strong baseline for future research on more capable VDR models.

\section*{Limitations}

In this work, we focus on VDR with text-only queries, leaving multimodal query formulations and downstream applications for future exploration. Another limitation is that SERVAL depends on VLM-generated text, which incurs an offline document preprocessing overhead and may introduce translation artifacts or hallucinations. These factors point to several promising directions for future work, including support for multimodal queries, hybrid text–layout representations, and systematic evaluation of hallucinations.

\section*{Ethical Considerations}

Our experiments rely on publicly released, pretrained vision–language models and text encoders. Prior work has shown that such models may encode social and linguistic biases \cite{hamidieh2024identifying, may2019measuring}. Because our study involves no additional training or fine-tuning, we do not introduce new biases; however, any biases present in the underlying models may still influence our results.

\section*{Acknowledgements}
This research was supported by project VI.Vidi.223.166 of the NWO Talent Programme which is (partly) financed by the Dutch Research Council NWO). 
We acknowledge the Dutch Research Council for awarding this project access to the LUMI supercomputer, owned by the EuroHPC Joint Undertaking, hosted by CSC (Finland) and the LUMI consortium through project number NWO-2024.050.

\bibliography{custom}

\appendix
\onecolumn
\section{Examples of generated descriptions by InternVL3}
\label{sec:generated_descriptions}

\begin{tcolorbox}[colback=white, colframe=blue!60!green, title=Bar Chart: Subchondral Bone Formation Comparison, sharp corners, boxrule=0.5pt]

\begin{wrapfigure}{l}{4cm}
  \vspace{-1em}
  \includegraphics[width=4cm,height=4cm]{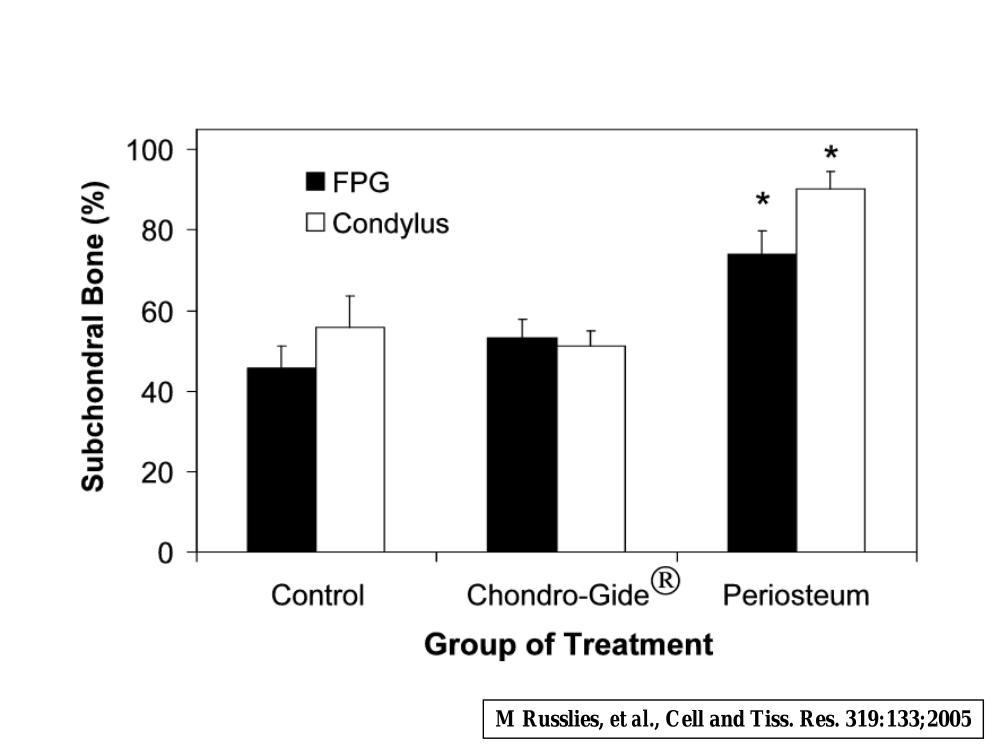}
\end{wrapfigure}

 \textbf{Summary}: The image displays a bar chart comparing the percentage of subchondral bone formation across different treatment groups. The chart illustrates the results for two conditions: FPG and Condylus. The groups compared are Control, Chondro-Gide® treated, and Periosteum treated. The results show that the Periosteum group has the highest subchondral bone percentage, with significant differences marked with an asterisk. \textbf{Details} Y-Axis: Labeled as "Subchondral Bone (\%)" ranging from 0 to 100. X-Axis: Labeled as "Group of Treatment," with categories: Control, Chondro-Gide®, and Periosteum. Bars: FPG (Filled Black Bars): Control: Approximately 40\% Chondro-Gide®: Approximately 55\% Periosteum: Approximately 75\% with an asterisk indicating significance. Condylus (Outlined White Bars): Control: Approximately 50\% Chondro-Gide®: Approximately 50\% Periosteum: Approximately 90\% with an asterisk indicating significance. \textbf{Visible Text and Numerical Values} Legend: FPG (filled square) Condylus (outlined square) X-Axis Labels: Control Chondro-Gide® Periosteum Y-Axis Labels: 0, 20, 40, 60, 80, 100 Asterisks: Indicating statistical significance for the Periosteum group in both conditions. Citation: "M Russlies, et al., Cell and Tiss. Res. 319:133;2005"

\vspace{1.5em}
\end{tcolorbox}

\begin{tcolorbox}[colback=white, colframe=blue!60!green, title=Complex Graph: Tesla vs S\&P 500 Stock Performance, sharp corners, boxrule=0.5pt]

\begin{wrapfigure}{l}{4cm}
  \vspace{-1em}
  \includegraphics[width=4cm,height=4cm]{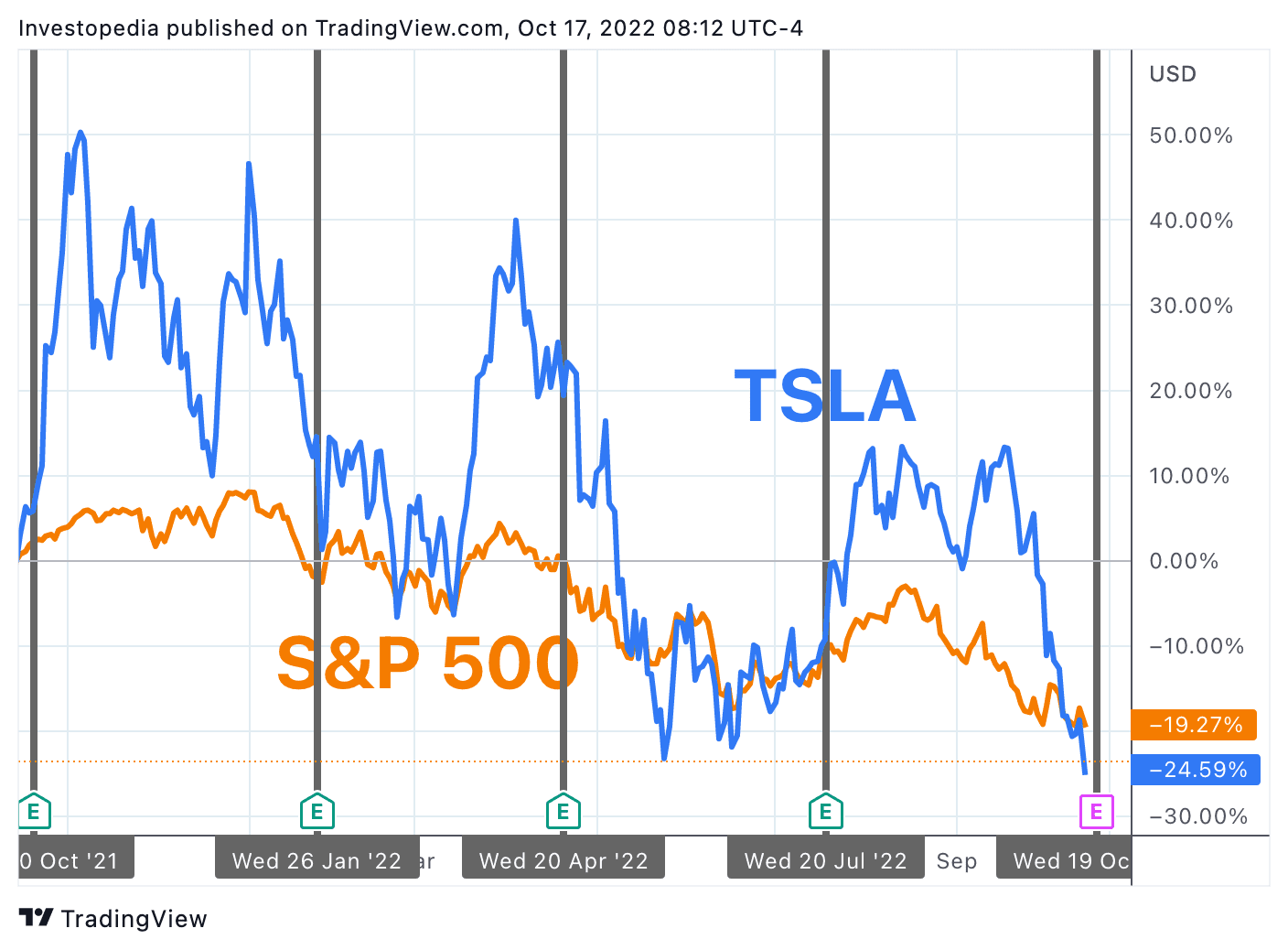}
\end{wrapfigure}
 The image is a financial chart comparing the performance of Tesla (TSLA) and the S \&P 500 over a specified period. It shows the percentage change in value over time, with Tesla represented by the blue line and the S \&P 500 by the orange line. Key earnings dates (marked with an "E") are highlighted on the chart.  \textbf{Summary}: The chart illustrates the relative performance of Tesla and the S \&P 500 from late 2021 to mid-2022. Tesla shows more volatility compared to the S \&P 500. Both indices experienced declines, with Tesla having a larger drop. \textbf{Details}: Tesla (TSLA, Blue Line): Starts around 0\% change. Peaks significantly multiple times, showing high volatility. Ends with a total decrease of approximately -24.59\%. S \&P 500 (Orange Line): Starts around 0\% change. Shows less volatility compared to Tesla. Ends with a decrease of approximately -19.27\%. Earnings Dates: Marked with an "E" on specific dates: Wed 26 Jan '22 Wed 20 Apr '22 Wed 20 Jul '22 Wed 19 Oct '22 Percentage Change Scale: Ranges from -30.00\% to 50.00\%. Chart Source: Published on TradingView.com by Investopedia on Oct 17, 2022, at 08:12 UTC-4. \textbf{Extracted Text and Numerical Values}: "Investopedia published on TradingView.com, Oct 17, 2022 08:12 UTC-4" "TSLA" "S \&P 500" Percentage change scale: "-30.00\%", "-20.00\%", "-10.00\%", "0.00\%", "10.00\%", "20.00\%", "30.00\%", "40.00\%", "50.00\%" Tesla change: "-24.59\%" S \&P 500 change: "-19.27\%" Dates: "0 Oct '21" "Wed 26 Jan '22" "Wed 20 Apr '22" "Wed 20 Jul '22" "Sep" "Wed 19 Oct" This chart provides a visual representation of how Tesla's stock has been more volatile and has underperformed relative to the S \&P 500 over the given period.

\vspace{1.5em}
\end{tcolorbox}

\begin{tcolorbox}[colback=white, colframe=blue!60!green, title=Complex Diagram: Guessing Game Flowchart, sharp corners, boxrule=0.5pt]

\begin{wrapfigure}{l}{4cm}
  \vspace{-1em}
  \includegraphics[width=4cm,height=3cm]{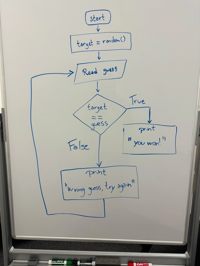}
\end{wrapfigure}

\textbf{Summary} The image shows a flowchart on a whiteboard that outlines a simple guessing game logic. The flowchart starts with generating a random target number and then prompts the user to guess the number. It checks if the guess matches the target and provides feedback accordingly, looping back if the guess is incorrect.  \textbf{Details} The flowchart begins with a "Start" step, followed by generating a random target number. The user is then prompted to input a guess. The guess is compared to the target number, and based on whether they match, different actions are taken:  If the guess is correct, it prints "you won!" and ends. If the guess is incorrect, it prints "wrong guess, try again" and loops back to reading another guess. \textbf{Extracted Text and Numerical Values} Start target = random() Read guess target == guess True print "you won!" False print "wrong guess, try again"

\vspace{1.5em}
\end{tcolorbox}

\begin{tcolorbox}[colback=white, colframe=blue!60!green, title=Visual Table: Macro Economics, sharp corners, boxrule=0.5pt]

\begin{wrapfigure}{l}{10cm}
  \vspace{-1em}
  \includegraphics[width=10cm,height=3cm]{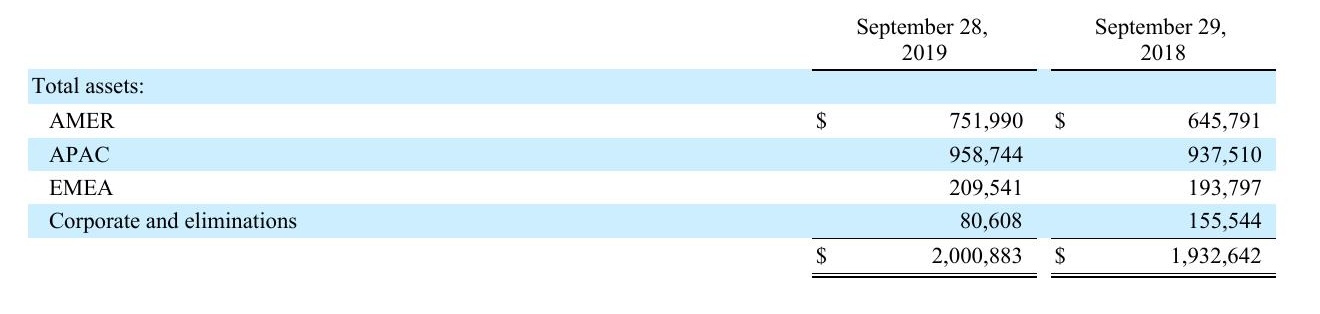}
\end{wrapfigure}

\textbf{Summary} The image displays a table summarizing the total assets for different regions and corporate eliminations as of September 28, 2019, and September 29, 2018. The regions listed are AMER, APAC, and EMEA. The table provides a comparative view of the total assets for each region and the corporate eliminations over the two years.  \textbf{Details} AMER September 28, 2019: \$751,990 September 29, 2018: \$645,791 APAC September 28, 2019: \$958,744 September 29, 2018: \$937,510 EMEA September 28, 2019: \$209,541 September 29, 2018: \$193,797 Corporate and eliminations September 28, 2019: \$80,608 September 29, 2018: \$155,544 Total Assets September 28, 2019: \$2,000,883 September 29, 2018: \$1,932,642 \textbf{Extracted Text and Numerical Values} Total assets: AMER September 28, 2019: \$751,990 September 29, 2018: \$645,791 APAC September 28, 2019: \$958,744 September 29, 2018: \$937,510 EMEA September 28, 2019: \$209,541 September 29, 2018: \$193,797 Corporate and eliminations September 28, 2019: \$80,608 September 29, 2018: \$155,544 Total September 28, 2019: \$2,000,883 September 29, 2018: \$1,932,642\$

\vspace{1.5em}
\end{tcolorbox}

\section{ViDoRe benchmark}
\label{sec:dataset_abbr}
The map from copora short names to its full name on HuggingFace is shown in Table \ref{tab:dataset-abbrev-map}.
\begin{table}[ht!]
\centering
\small
\begin{tabular}{ll}
\toprule
\toprule
\textbf{Abbreviation} & \textbf{HuggingFace Path} \\ \midrule
\textbf{RERB}   & \texttt{vidore/restaurant\_esg\_reports\_beir} \\
\textbf{SAXA}   & \texttt{vidore/synthetic\_axa\_filtered\_v1.0} \\
\textbf{SAXAM}  & \texttt{vidore/synthetic\_axa\_filtered\_v1.0\_multilingual} \\
\textbf{SEME}   & \texttt{vidore/synthetic\_economics\_macro\_economy\_2024\_filtered\_v1.0} \\
\textbf{SMBTI}  & \texttt{vidore/synthetic\_mit\_biomedical\_tissue\_interactions\_unfiltered} \\
\textbf{SMBTIM} & \texttt{vidore/synthetic\_mit\_biomedical\_tissue\_interactions\_unfiltered\_multilingual} \\
\textbf{SRS}    & \texttt{vidore/synthetic\_rse\_restaurant\_filtered\_v1.0} \\
\textbf{SRSM}   & \texttt{vidore/synthetic\_rse\_restaurant\_filtered\_v1.0\_multilingual} \\
\textbf{SEMEM}  & \texttt{vidore/synthetics\_economics\_macro\_economy\_2024\_filtered\_v1.0\_multilingual} \\
\bottomrule
\bottomrule
\end{tabular}
\caption{Mapping from dataset abbreviations to their full paths.}
\label{tab:dataset-abbrev-map}
\end{table}

\section{Detailed Zeroshot VDR Results}

The full evaluation results, including all metrics (nDCG@1, nDCG@5, nDCG@10, R@1, R@5, R@10) are shown in Table \ref{tab:results_ndcg1}, Table \ref{tab:results_ndcg5}, Table \ref{tab:results_ndcg10}, Table \ref{tab:results_r1}, Table \ref{tab:results_r5}, Table \ref{tab:results_r10} respectively.
\label{sec:detailed_zeroshot_results}
\begin{table*}[ht!]
  \centering
  \tiny
  \setlength{\tabcolsep}{3pt}
  \begin{tabular}{l l c c c c c c c c c | c}
\toprule
\toprule
\textbf{VLM} & \textbf{Text Encoder} & \textbf{RERB} & \textbf{SAXA} & \textbf{SAXAM} & \textbf{SEME} & \textbf{SMBTI} & \textbf{SMBTIM} & \textbf{SRS} & \textbf{SRSM} & \textbf{SEMEM} & \textbf{AVG} \\
\midrule
\multirow{7}{*}{Qwen2.5VL-7B} & splade 108M~\cite{lassance2024splade} & 61.2 & 33.3 & 30.6 & 60.3 & 57.5 & 36.2 & 33.3 & 30.3 & 28.4 & 41.2 \\
 & open search v1 133M~\cite{geng2024towards} & 60.6 & 22.2 & 30.6 & 63.8 & 53.8 & 36.7 & 29.8 & 27.2 & 31.0 & 39.5 \\
 & me5\_base 278M~\cite{wang2024multilingual} & 36.5 & 44.4 & 50.0 & 60.3 & 49.4 & 44.4 & 40.4 & 41.7 & 47.0 & 46.0 \\
 & me5\_large 560M~\cite{wang2024multilingual} & 58.3 & 44.4 & 56.9 & 56.9 & 54.4 & 49.8 & 54.4 & 55.3 & 50.9 & 53.5 \\
 & inf\_small 1.5B~\cite{infly-ai_2025} & 58.7 & 61.1 & 66.7 & 55.2 & 58.8 & 51.7 & 56.1 & 50.9 & 51.3 & 56.7 \\
 & inf 7B~\cite{infly-ai_2025} & 63.5 & 72.2 & 70.8 & 55.2 & 61.3 & 59.5 & 59.6 & 57.0 & 54.3 & 61.5 \\
 & sfr 7B~\cite{infly-ai_2025}  & 63.5 & 72.2 & 69.4 & 55.2 & 56.9 & 53.1 & 57.9 & 56.6 & 48.3 & 59.2 \\
\midrule
\multirow{7}{*}{Qwen2.5VL-32B} & splade 108M~\cite{lassance2024splade} & 66.0 & 55.6 & 36.1 & 70.7 & 62.5 & 37.8 & 29.8 & 30.7 & 34.5 & 47.1 \\
 & open search v1 133M~\cite{geng2024towards} & 59.6 & 27.8 & 33.3 & 72.4 & 62.5 & 36.9 & 28.1 & 28.1 & 34.5 & 42.6 \\
 & me5\_base 278M~\cite{wang2024multilingual} & 54.5 & 50.0 & 55.6 & 56.9 & 50.6 & 42.5 & 50.9 & 46.5 & 45.7 & 50.3 \\
 & me5\_large 560M~\cite{wang2024multilingual} & 64.1 & 50.0 & 51.4 & 53.4 & 56.9 & 53.4 & 52.6 & 55.7 & 52.2 & 54.4 \\
 & inf\_small 1.5B~\cite{infly-ai_2025} & 60.3 & 77.8 & 70.8 & 56.9 & 58.8 & 55.3 & 45.6 & 48.7 & 52.2 & 58.5 \\
 & inf 7B~\cite{infly-ai_2025} & 67.9 & 72.2 & 76.4 & 58.6 & 60.6 & 59.8 & 56.1 & 57.9 & 54.3 & 62.7 \\
 & sfr 7B~\cite{infly-ai_2025}  & 58.3 & 72.2 & 75.0 & 60.3 & 61.9 & 59.1 & 56.1 & 56.1 & 53.4 & 61.4 \\
\midrule
\multirow{7}{*}{Qwen2.5VL-72B} & splade 108M~\cite{lassance2024splade} & 56.4 & 16.7 & 25.0 & 65.5 & 60.6 & 39.8 & 28.1 & 31.1 & 31.9 & 39.5 \\
 & open search v1 133M~\cite{geng2024towards} & 58.3 & 16.7 & 22.2 & 70.7 & 58.1 & 39.5 & 29.8 & 28.5 & 27.2 & 39.0 \\
 & me5\_base 278M~\cite{wang2024multilingual} & 32.7 & 44.4 & 59.7 & 53.4 & 54.4 & 45.0 & 47.4 & 44.3 & 46.1 & 47.5 \\
 & me5\_large 560M~\cite{wang2024multilingual} & 59.6 & 44.4 & 55.6 & 56.9 & 58.8 & 53.1 & 50.9 & 47.8 & 53.4 & 53.4 \\
 & inf\_small 1.5B~\cite{infly-ai_2025} & 58.3 & 50.0 & 62.5 & 69.0 & 60.0 & 53.0 & 43.9 & 46.9 & 54.3 & 55.3 \\
 & inf 7B~\cite{infly-ai_2025} & 67.9 & 72.2 & 69.4 & 62.1 & 63.1 & 60.5 & 56.1 & 53.9 & 62.9 & 63.1 \\
 & sfr 7B~\cite{infly-ai_2025}  & 62.2 & 61.1 & 66.7 & 58.6 & 60.0 & 55.9 & 56.1 & 53.9 & 49.1 & 58.2 \\
\midrule
\multirow{7}{*}{InternVL3-2B} & splade 108M~\cite{lassance2024splade} & 50.6 & 50.0 & 29.2 & 75.9 & 53.8 & 35.5 & 21.1 & 25.9 & 33.6 & 41.7 \\
 & open search v1 133M~\cite{geng2024towards} & 41.0 & 61.1 & 33.3 & 74.1 & 53.1 & 34.2 & 19.3 & 24.1 & 31.9 & 41.4 \\
 & me5\_base 278M~\cite{wang2024multilingual} & 39.1 & 55.6 & 38.9 & 46.6 & 41.9 & 39.4 & 35.1 & 31.6 & 40.5 & 40.9 \\
 & me5\_large 560M~\cite{wang2024multilingual} & 55.8 & 44.4 & 43.1 & 60.3 & 49.4 & 48.4 & 52.6 & 48.2 & 45.7 & 49.8 \\
 & inf\_small 1.5B~\cite{infly-ai_2025} & 53.8 & 66.7 & 58.3 & 60.3 & 55.0 & 49.8 & 33.3 & 37.3 & 47.4 & 51.3 \\
 & inf 7B~\cite{infly-ai_2025} & 61.5 & 83.3 & 66.7 & 56.9 & 59.4 & 57.2 & 45.6 & 44.7 & 53.0 & 58.7 \\
 & sfr 7B~\cite{infly-ai_2025}  & 59.6 & 66.7 & 51.4 & 51.7 & 59.4 & 57.0 & 56.1 & 54.4 & 41.8 & 55.3 \\
\midrule
\multirow{7}{*}{InternVL3-8B} & splade 108M~\cite{lassance2024splade} & 55.1 & 27.8 & 27.8 & 67.2 & 59.4 & 37.5 & 28.1 & 32.5 & 32.3 & 40.9 \\
 & open search v1 133M~\cite{geng2024towards} & 52.6 & 33.3 & 34.7 & 62.1 & 55.6 & 36.9 & 35.1 & 32.5 & 29.7 & 41.4 \\
 & me5\_base 278M~\cite{wang2024multilingual} & 46.8 & 55.6 & 50.0 & 55.2 & 50.0 & 46.7 & 38.6 & 41.7 & 47.4 & 48.0 \\
 & me5\_large 560M~\cite{wang2024multilingual} & 46.8 & 16.7 & 43.1 & 58.6 & 55.6 & 54.5 & 49.1 & 51.3 & 51.3 & 47.4 \\
 & inf\_small 1.5B~\cite{infly-ai_2025} & 59.9 & 61.1 & 55.6 & 69.0 & 58.1 & 50.2 & 47.4 & 48.7 & 59.9 & 56.6 \\
 & inf 7B~\cite{infly-ai_2025} & 65.4 & 72.2 & 70.8 & 67.2 & 60.6 & 59.8 & 57.9 & 54.8 & 63.4 & 63.6 \\
 & sfr 7B~\cite{infly-ai_2025}  & 55.1 & 44.4 & 61.1 & 55.2 & 56.9 & 56.4 & 54.4 & 54.8 & 49.6 & 54.2 \\
\midrule
\multirow{7}{*}{InternVL3-14B} & splade 108M~\cite{lassance2024splade} & 62.2 & 33.3 & 27.8 & 51.7 & 60.0 & 37.0 & 35.1 & 36.0 & 24.1 & 40.8 \\
 & open search v1 133M~\cite{geng2024towards} & 54.5 & 33.3 & 27.8 & 51.7 & 60.0 & 37.8 & 24.6 & 29.8 & 25.0 & 38.3 \\
 & me5\_base 278M~\cite{wang2024multilingual} & 35.3 & 55.6 & 48.6 & 53.4 & 48.1 & 41.6 & 43.9 & 39.9 & 40.9 & 45.3 \\
 & me5\_large 560M~\cite{wang2024multilingual} & 63.5 & 38.9 & 44.4 & 55.2 & 55.0 & 51.6 & 49.1 & 51.3 & 48.3 & 50.8 \\
 & inf\_small 1.5B~\cite{infly-ai_2025} & 56.4 & 61.1 & 56.9 & 62.1 & 58.1 & 50.2 & 43.9 & 43.4 & 55.6 & 54.2 \\
 & inf 7B~\cite{infly-ai_2025} & 61.5 & 72.2 & 69.4 & 60.3 & 64.4 & 60.2 & 50.9 & 51.3 & 61.2 & 61.3 \\
 & sfr 7B~\cite{infly-ai_2025}  & 64.1 & 72.2 & 58.3 & 48.3 & 55.0 & 55.0 & 57.9 & 59.2 & 47.8 & 57.5 \\
\bottomrule
\bottomrule
\end{tabular}
  \caption{Zeroshot VDR performance on ViDoRe-v2 benchmark. Metric: nDCG@1.}
  \label{tab:results_ndcg1}
  
\end{table*}

\begin{table*}
  \centering
  \tiny
  \setlength{\tabcolsep}{3pt}
  \begin{tabular}{l l c c c c c c c c c | c}
\toprule
\toprule
\textbf{VLM} & \textbf{Text Encoder} & \textbf{RERB} & \textbf{SAXA} & \textbf{SAXAM} & \textbf{SEME} & \textbf{SMBTI} & \textbf{SMBTIM} & \textbf{SRS} & \textbf{SRSM} & \textbf{SEMEM} & \textbf{AVG} \\
\midrule
\multirow{7}{*}{Qwen2.5VL-7B} & splade 108M~\cite{lassance2024splade} & 64.8 & 32.1 & 30.6 & 56.4 & 59.8 & 39.8 & 33.8 & 35.4 & 27.4 & 42.2 \\
 & open search v1 133M~\cite{geng2024towards} & 61.0 & 29.0 & 30.5 & 59.1 & 59.7 & 41.3 & 33.7 & 34.8 & 29.5 & 42.1 \\
 & me5\_base 278M~\cite{wang2024multilingual} & 45.2 & 48.3 & 51.7 & 49.1 & 53.9 & 48.2 & 44.1 & 44.9 & 40.7 & 47.3 \\
 & me5\_large 560M~\cite{wang2024multilingual} & 64.5 & 56.5 & 58.0 & 52.5 & 58.6 & 55.9 & 55.9 & 55.5 & 47.8 & 56.1 \\
 & inf\_small 1.5B~\cite{infly-ai_2025} & 60.2 & 65.2 & 64.4 & 58.1 & 60.6 & 57.1 & 57.1 & 54.9 & 51.5 & 58.8 \\
 & inf 7B~\cite{infly-ai_2025} & 66.3 & 72.0 & 65.0 & 57.2 & 64.9 & 62.2 & 57.4 & 56.3 & 54.5 & 61.8 \\
 & sfr 7B~\cite{infly-ai_2025}  & 65.8 & 70.3 & 65.7 & 53.6 & 59.8 & 58.0 & 56.3 & 57.9 & 47.5 & 59.4 \\
\midrule
\multirow{7}{*}{Qwen2.5VL-32B} & splade 108M~\cite{lassance2024splade} & 69.3 & 53.1 & 38.5 & 58.1 & 63.1 & 40.8 & 35.0 & 35.2 & 29.4 & 46.9 \\
 & open search v1 133M~\cite{geng2024towards} & 68.0 & 45.6 & 39.1 & 59.7 & 63.0 & 41.4 & 34.2 & 33.8 & 28.8 & 46.0 \\
 & me5\_base 278M~\cite{wang2024multilingual} & 59.0 & 49.9 & 52.2 & 50.0 & 55.5 & 48.8 & 51.3 & 51.9 & 44.2 & 51.4 \\
 & me5\_large 560M~\cite{wang2024multilingual} & 68.8 & 53.8 & 53.5 & 49.1 & 60.6 & 57.1 & 52.0 & 54.0 & 46.2 & 55.0 \\
 & inf\_small 1.5B~\cite{infly-ai_2025} & 67.9 & 68.3 & 66.4 & 56.6 & 60.9 & 58.6 & 54.8 & 55.3 & 50.0 & 59.9 \\
 & inf 7B~\cite{infly-ai_2025} & 70.7 & 69.7 & 69.4 & 59.0 & 65.1 & 63.0 & 59.0 & 59.5 & 55.5 & 63.4 \\
 & sfr 7B~\cite{infly-ai_2025}  & 65.7 & 72.8 & 72.1 & 54.4 & 61.5 & 59.6 & 57.9 & 59.1 & 47.9 & 61.2 \\
\midrule
\multirow{7}{*}{Qwen2.5VL-72B} & splade 108M~\cite{lassance2024splade} & 63.8 & 16.4 & 25.8 & 59.8 & 62.7 & 42.1 & 31.9 & 37.0 & 30.1 & 41.1 \\
 & open search v1 133M~\cite{geng2024towards} & 61.5 & 20.2 & 28.1 & 62.4 & 61.2 & 42.4 & 31.9 & 35.3 & 28.9 & 41.3 \\
 & me5\_base 278M~\cite{wang2024multilingual} & 45.6 & 49.5 & 55.3 & 49.6 & 57.4 & 48.7 & 43.7 & 46.1 & 42.9 & 48.8 \\
 & me5\_large 560M~\cite{wang2024multilingual} & 66.1 & 50.8 & 56.0 & 52.5 & 61.6 & 58.0 & 50.5 & 49.0 & 49.2 & 54.9 \\
 & inf\_small 1.5B~\cite{infly-ai_2025} & 67.3 & 58.8 & 62.1 & 62.7 & 62.6 & 59.3 & 51.7 & 53.7 & 52.6 & 59.0 \\
 & inf 7B~\cite{infly-ai_2025} & 68.9 & 70.7 & 67.9 & 60.3 & 65.5 & 63.2 & 58.2 & 58.6 & 57.7 & 63.4 \\
 & sfr 7B~\cite{infly-ai_2025}  & 66.0 & 66.0 & 66.8 & 54.4 & 62.2 & 60.4 & 59.0 & 58.1 & 46.9 & 60.0 \\
\midrule
\multirow{7}{*}{InternVL3-2B} & splade 108M~\cite{lassance2024splade} & 52.7 & 48.1 & 28.6 & 58.6 & 58.6 & 39.6 & 28.4 & 33.5 & 30.0 & 42.0 \\
 & open search v1 133M~\cite{geng2024towards} & 48.6 & 59.4 & 32.4 & 59.9 & 58.7 & 39.3 & 29.1 & 31.2 & 29.9 & 43.2 \\
 & me5\_base 278M~\cite{wang2024multilingual} & 44.0 & 52.0 & 36.5 & 47.1 & 50.4 & 46.3 & 38.6 & 38.1 & 39.8 & 43.6 \\
 & me5\_large 560M~\cite{wang2024multilingual} & 61.4 & 52.9 & 42.0 & 51.4 & 56.7 & 54.8 & 52.6 & 50.7 & 43.9 & 51.8 \\
 & inf\_small 1.5B~\cite{infly-ai_2025} & 59.8 & 62.5 & 54.1 & 52.9 & 59.3 & 55.2 & 47.2 & 48.7 & 46.2 & 54.0 \\
 & inf 7B~\cite{infly-ai_2025} & 63.3 & 72.5 & 62.6 & 55.0 & 63.8 & 60.7 & 52.9 & 52.1 & 51.3 & 59.4 \\
 & sfr 7B~\cite{infly-ai_2025}  & 64.5 & 68.0 & 55.8 & 45.7 & 62.5 & 59.9 & 54.8 & 55.2 & 41.3 & 56.4 \\
\midrule
\multirow{7}{*}{InternVL3-8B} & splade 108M~\cite{lassance2024splade} & 65.6 & 30.7 & 31.2 & 58.0 & 60.6 & 40.6 & 34.9 & 38.2 & 30.3 & 43.3 \\
 & open search v1 133M~\cite{geng2024towards} & 60.2 & 32.0 & 34.0 & 59.0 & 59.5 & 41.0 & 36.4 & 37.9 & 29.5 & 43.3 \\
 & me5\_base 278M~\cite{wang2024multilingual} & 52.0 & 48.7 & 50.7 & 48.8 & 52.7 & 49.1 & 41.9 & 45.3 & 41.3 & 47.8 \\
 & me5\_large 560M~\cite{wang2024multilingual} & 58.1 & 38.8 & 49.6 & 52.2 & 59.8 & 57.7 & 53.1 & 56.0 & 47.5 & 52.5 \\
 & inf\_small 1.5B~\cite{infly-ai_2025} & 63.0 & 54.4 & 57.6 & 58.8 & 63.4 & 57.8 & 50.6 & 51.9 & 52.5 & 56.7 \\
 & inf 7B~\cite{infly-ai_2025} & 66.7 & 62.9 & 64.8 & 57.3 & 65.5 & 63.0 & 57.6 & 57.4 & 54.6 & 61.1 \\
 & sfr 7B~\cite{infly-ai_2025}  & 63.9 & 57.2 & 62.0 & 50.0 & 62.0 & 59.5 & 57.5 & 57.2 & 44.9 & 57.1 \\
\midrule
\multirow{7}{*}{InternVL3-14B} & splade 108M~\cite{lassance2024splade} & 68.5 & 38.0 & 31.4 & 53.8 & 61.0 & 40.7 & 36.9 & 37.6 & 26.1 & 43.8 \\
 & open search v1 133M~\cite{geng2024towards} & 61.3 & 39.8 & 32.7 & 54.7 & 60.3 & 41.1 & 32.2 & 34.8 & 27.0 & 42.7 \\
 & me5\_base 278M~\cite{wang2024multilingual} & 47.7 & 56.2 & 52.9 & 49.7 & 51.7 & 46.3 & 45.2 & 45.0 & 40.0 & 48.3 \\
 & me5\_large 560M~\cite{wang2024multilingual} & 67.7 & 54.3 & 53.0 & 52.4 & 58.3 & 55.2 & 54.2 & 55.1 & 44.9 & 55.0 \\
 & inf\_small 1.5B~\cite{infly-ai_2025} & 60.5 & 60.1 & 57.3 & 58.0 & 62.4 & 57.6 & 51.2 & 52.1 & 51.7 & 56.8 \\
 & inf 7B~\cite{infly-ai_2025} & 67.7 & 64.7 & 63.0 & 58.9 & 64.7 & 62.7 & 52.4 & 54.6 & 56.5 & 60.6 \\
 & sfr 7B~\cite{infly-ai_2025}  & 66.8 & 69.1 & 61.5 & 49.8 & 61.0 & 59.8 & 55.0 & 57.0 & 45.0 & 58.3 \\

\bottomrule
\bottomrule
\end{tabular}
  \caption{Zeroshot VDR performance on ViDoRe-v2 benchmark. Metric: nDCG@5.}
  \label{tab:results_ndcg5}
  
\end{table*}
\begin{table*}
  \centering
  \tiny 
  \setlength{\tabcolsep}{3pt}
  \begin{tabular}{l l c c c c c c c c c | c}
\toprule
\toprule
\textbf{VLM} & \textbf{Text Encoder} & \textbf{RERB} & \textbf{SAXA} & \textbf{SAXAM} & \textbf{SEME} & \textbf{SMBTI} & \textbf{SMBTIM} & \textbf{SRS} & \textbf{SRSM} & \textbf{SEMEM} & \textbf{AVG} \\
\midrule
\multirow{7}{*}{Qwen2.5VL-7B} & splade 108M~\cite{lassance2024splade} & 68.0 & 32.3 & 31.4 & 56.4 & 63.3 & 43.3 & 40.1 & 41.0 & 28.8 & 44.9 \\
 & open search v1 133M~\cite{geng2024towards} & 65.4 & 30.9 & 31.5 & 55.6 & 62.5 & 44.1 & 37.9 & 38.4 & 28.9 & 43.9 \\
 & me5\_base 278M~\cite{wang2024multilingual} & 50.1 & 49.5 & 51.6 & 48.7 & 56.7 & 51.4 & 50.4 & 50.2 & 40.1 & 49.9 \\
 & me5\_large 560M~\cite{wang2024multilingual} & 68.1 & 56.6 & 58.1 & 50.1 & 63.6 & 60.1 & 59.5 & 59.6 & 46.1 & 58.0 \\
 & inf\_small 1.5B~\cite{infly-ai_2025} & 65.4 & 63.6 & 64.3 & 54.6 & 65.2 & 61.0 & 60.5 & 59.3 & 50.0 & 60.4 \\
 & inf 7B~\cite{infly-ai_2025} & 68.4 & 70.1 & 65.8 & 54.9 & 68.0 & 65.5 & 60.9 & 59.5 & 53.1 & 62.9 \\
 & sfr 7B~\cite{infly-ai_2025}  & 68.1 & 69.6 & 67.4 & 50.8 & 63.2 & 61.3 & 60.7 & 62.4 & 46.7 & 61.1 \\
\midrule
\multirow{7}{*}{Qwen2.5VL-32B} & splade 108M~\cite{lassance2024splade} & 74.2 & 54.7 & 41.8 & 57.4 & 66.0 & 44.4 & 41.0 & 41.0 & 30.0 & 50.1 \\
 & open search v1 133M~\cite{geng2024towards} & 71.7 & 50.5 & 42.8 & 58.8 & 65.8 & 45.1 & 40.6 & 39.3 & 30.0 & 49.4 \\
 & me5\_base 278M~\cite{wang2024multilingual} & 61.6 & 51.2 & 53.8 & 47.5 & 59.5 & 53.1 & 56.3 & 55.4 & 41.6 & 53.3 \\
 & me5\_large 560M~\cite{wang2024multilingual} & 71.5 & 55.2 & 55.0 & 47.6 & 64.2 & 60.7 & 56.8 & 57.9 & 45.0 & 57.1 \\
 & inf\_small 1.5B~\cite{infly-ai_2025} & 69.6 & 68.3 & 65.6 & 55.7 & 64.7 & 61.9 & 58.2 & 60.0 & 49.0 & 61.5 \\
 & inf 7B~\cite{infly-ai_2025} & 73.9 & 69.9 & 69.2 & 55.4 & 68.0 & 66.3 & 64.5 & 64.7 & 52.9 & 65.0 \\
 & sfr 7B~\cite{infly-ai_2025}  & 68.0 & 70.4 & 71.3 & 51.2 & 65.0 & 63.1 & 62.5 & 63.4 & 46.9 & 62.4 \\
\midrule
\multirow{7}{*}{Qwen2.5VL-72B} & splade 108M~\cite{lassance2024splade} & 68.1 & 19.3 & 28.8 & 56.2 & 66.0 & 45.4 & 38.3 & 42.0 & 30.7 & 43.9 \\
 & open search v1 133M~\cite{geng2024towards} & 66.5 & 24.5 & 31.1 & 59.3 & 64.7 & 45.4 & 39.5 & 40.1 & 29.6 & 44.5 \\
 & me5\_base 278M~\cite{wang2024multilingual} & 48.5 & 51.7 & 56.4 & 49.0 & 61.0 & 52.8 & 49.6 & 50.9 & 41.7 & 51.3 \\
 & me5\_large 560M~\cite{wang2024multilingual} & 68.1 & 52.6 & 57.1 & 51.6 & 64.9 & 61.6 & 55.1 & 54.2 & 47.9 & 57.0 \\
 & inf\_small 1.5B~\cite{infly-ai_2025} & 69.0 & 59.2 & 61.5 & 59.5 & 65.8 & 62.4 & 56.6 & 59.3 & 51.2 & 60.5 \\
 & inf 7B~\cite{infly-ai_2025} & 70.8 & 70.8 & 66.9 & 58.2 & 68.5 & 66.8 & 63.4 & 63.6 & 56.1 & 65.0 \\
 & sfr 7B~\cite{infly-ai_2025}  & 70.2 & 66.6 & 66.5 & 52.6 & 67.1 & 63.7 & 61.9 & 62.9 & 46.8 & 62.0 \\
\midrule
\multirow{7}{*}{InternVL3-2B} & splade 108M~\cite{lassance2024splade} & 59.0 & 51.5 & 29.6 & 58.6 & 62.8 & 43.3 & 35.6 & 39.9 & 30.7 & 45.7 \\
 & open search v1 133M~\cite{geng2024towards} & 54.2 & 60.0 & 33.1 & 58.4 & 62.4 & 43.1 & 35.0 & 37.2 & 30.4 & 46.0 \\
 & me5\_base 278M~\cite{wang2024multilingual} & 47.7 & 53.6 & 39.2 & 46.0 & 54.7 & 50.0 & 42.4 & 42.3 & 39.3 & 46.1 \\
 & me5\_large 560M~\cite{wang2024multilingual} & 65.5 & 55.4 & 45.5 & 48.4 & 60.6 & 58.4 & 54.1 & 53.4 & 43.0 & 53.8 \\
 & inf\_small 1.5B~\cite{infly-ai_2025} & 61.7 & 61.4 & 54.9 & 53.4 & 63.1 & 58.6 & 51.8 & 53.7 & 47.1 & 56.2 \\
 & inf 7B~\cite{infly-ai_2025} & 68.3 & 72.7 & 63.7 & 54.2 & 67.2 & 64.4 & 56.6 & 55.8 & 51.2 & 61.6 \\
 & sfr 7B~\cite{infly-ai_2025}  & 66.4 & 65.3 & 55.5 & 43.9 & 66.1 & 63.3 & 59.1 & 59.9 & 41.2 & 57.9 \\
\midrule
\multirow{7}{*}{InternVL3-8B} & splade 108M~\cite{lassance2024splade} & 67.3 & 32.0 & 33.3 & 56.0 & 64.1 & 44.1 & 40.1 & 43.5 & 31.0 & 45.7 \\
 & open search v1 133M~\cite{geng2024towards} & 65.0 & 37.3 & 36.1 & 55.3 & 63.0 & 44.4 & 42.2 & 43.7 & 29.3 & 46.3 \\
 & me5\_base 278M~\cite{wang2024multilingual} & 57.6 & 49.5 & 51.0 & 47.1 & 57.2 & 52.7 & 48.9 & 50.4 & 40.2 & 50.5 \\
 & me5\_large 560M~\cite{wang2024multilingual} & 62.9 & 41.0 & 51.8 & 49.0 & 63.3 & 60.9 & 58.5 & 60.5 & 45.3 & 54.8 \\
 & inf\_small 1.5B~\cite{infly-ai_2025} & 66.0 & 54.4 & 58.2 & 57.3 & 65.9 & 60.7 & 56.2 & 57.3 & 51.2 & 58.6 \\
 & inf 7B~\cite{infly-ai_2025} & 70.5 & 65.4 & 64.0 & 56.4 & 68.4 & 66.5 & 61.6 & 61.7 & 53.6 & 63.1 \\
 & sfr 7B~\cite{infly-ai_2025}  & 68.1 & 58.8 & 63.0 & 48.0 & 64.6 & 63.4 & 61.1 & 62.1 & 43.9 & 59.2 \\
\midrule
\multirow{7}{*}{InternVL3-14B} & splade 108M~\cite{lassance2024splade} & 70.6 & 41.8 & 34.0 & 53.4 & 64.5 & 44.2 & 41.8 & 42.8 & 28.0 & 46.8 \\
 & open search v1 133M~\cite{geng2024towards} & 64.4 & 45.6 & 35.9 & 55.0 & 63.9 & 44.7 & 37.6 & 40.4 & 28.6 & 46.2 \\
 & me5\_base 278M~\cite{wang2024multilingual} & 53.4 & 58.8 & 56.1 & 46.6 & 55.4 & 50.1 & 49.2 & 49.3 & 39.7 & 50.9 \\
 & me5\_large 560M~\cite{wang2024multilingual} & 70.8 & 56.1 & 54.6 & 50.3 & 62.1 & 59.0 & 58.3 & 59.1 & 44.1 & 57.2 \\
 & inf\_small 1.5B~\cite{infly-ai_2025} & 63.1 & 61.2 & 58.9 & 57.3 & 66.0 & 61.4 & 56.0 & 56.7 & 50.9 & 59.1 \\
 & inf 7B~\cite{infly-ai_2025} & 70.1 & 65.5 & 64.7 & 56.0 & 68.0 & 66.2 & 58.1 & 59.5 & 54.2 & 62.5 \\
 & sfr 7B~\cite{infly-ai_2025}  & 69.0 & 67.1 & 62.9 & 48.1 & 64.3 & 63.0 & 60.3 & 62.5 & 45.0 & 60.3 \\
\bottomrule
\bottomrule
\end{tabular}
  \caption{Zeroshot VDR performance on ViDoRe-v2 benchmark. Metric: nDCG@10.}
  \label{tab:results_ndcg10}
  
\end{table*}

\begin{table*}
  \centering
  \tiny 
  \setlength{\tabcolsep}{3pt}
  \begin{tabular}{l l c c c c c c c c c | c}
\toprule
\toprule
\textbf{VLM} & \textbf{Text Encoder} & \textbf{RERB} & \textbf{SAXA} & \textbf{SAXAM} & \textbf{SEME} & \textbf{SMBTI} & \textbf{SMBTIM} & \textbf{SRS} & \textbf{SRSM} & \textbf{SEMEM} & \textbf{AVG} \\

\midrule
\multirow{7}{*}{Qwen2.5VL-7B} & splade 108M~\cite{lassance2024splade} & 41.6 & 12.0 & 12.6 & 10.0 & 33.7 & 21.9 & 15.5 & 14.6 & 3.7 & 18.4 \\
 & open search v1 133M~\cite{geng2024towards} & 40.8 & 10.6 & 12.7 & 11.0 & 33.3 & 22.3 & 13.7 & 12.9 & 4.2 & 18.0 \\
 & me5\_base 278M~\cite{wang2024multilingual} & 24.3 & 23.1 & 26.2 & 6.9 & 30.3 & 26.1 & 20.9 & 20.7 & 4.8 & 20.4 \\
 & me5\_large 560M~\cite{wang2024multilingual} & 44.3 & 24.1 & 27.7 & 7.6 & 32.9 & 30.2 & 27.6 & 27.8 & 7.5 & 25.5 \\
 & inf\_small 1.5B~\cite{infly-ai_2025} & 41.3 & 25.6 & 31.0 & 5.9 & 36.1 & 32.5 & 27.7 & 24.1 & 7.5 & 25.8 \\
 & inf 7B~\cite{infly-ai_2025} & 45.1 & 36.8 & 35.2 & 5.8 & 38.5 & 36.5 & 26.5 & 25.3 & 8.2 & 28.7 \\
 & sfr 7B~\cite{infly-ai_2025}  & 43.7 & 31.8 & 33.6 & 7.1 & 32.0 & 29.8 & 25.2 & 25.1 & 6.1 & 26.0 \\
\midrule
\multirow{7}{*}{Qwen2.5VL-32B} & splade 108M~\cite{lassance2024splade} & 45.0 & 28.5 & 18.0 & 11.1 & 36.6 & 23.7 & 13.8 & 13.8 & 4.3 & 21.6 \\
 & open search v1 133M~\cite{geng2024towards} & 42.1 & 16.2 & 17.7 & 11.4 & 36.7 & 22.9 & 14.2 & 13.5 & 4.4 & 19.9 \\
 & me5\_base 278M~\cite{wang2024multilingual} & 38.5 & 29.6 & 27.4 & 6.5 & 32.3 & 27.1 & 21.9 & 20.4 & 5.3 & 23.2 \\
 & me5\_large 560M~\cite{wang2024multilingual} & 44.8 & 28.7 & 30.1 & 6.7 & 34.5 & 31.7 & 25.4 & 27.1 & 6.0 & 26.1 \\
 & inf\_small 1.5B~\cite{infly-ai_2025} & 43.2 & 37.4 & 34.9 & 6.3 & 34.6 & 33.7 & 23.6 & 25.3 & 5.0 & 27.1 \\
 & inf 7B~\cite{infly-ai_2025} & 51.5 & 36.8 & 39.5 & 7.3 & 37.6 & 36.9 & 28.7 & 28.8 & 6.7 & 30.4 \\
 & sfr 7B~\cite{infly-ai_2025}  & 42.0 & 36.7 & 39.3 & 10.3 & 35.3 & 34.2 & 24.9 & 24.7 & 9.2 & 28.5 \\
\midrule
\multirow{7}{*}{Qwen2.5VL-72B} & splade 108M~\cite{lassance2024splade} & 37.5 & 2.2 & 9.0 & 8.8 & 36.8 & 24.8 & 14.3 & 14.0 & 3.9 & 16.8 \\
 & open search v1 133M~\cite{geng2024towards} & 39.5 & 2.7 & 7.8 & 9.7 & 35.3 & 24.7 & 13.1 & 12.9 & 2.8 & 16.5 \\
 & me5\_base 278M~\cite{wang2024multilingual} & 19.8 & 23.8 & 30.2 & 5.3 & 34.3 & 27.2 & 19.4 & 17.7 & 5.1 & 20.3 \\
 & me5\_large 560M~\cite{wang2024multilingual} & 41.8 & 20.7 & 26.2 & 9.4 & 34.6 & 30.9 & 24.1 & 21.4 & 9.9 & 24.3 \\
 & inf\_small 1.5B~\cite{infly-ai_2025} & 42.5 & 21.9 & 28.6 & 13.4 & 37.8 & 33.2 & 21.5 & 23.2 & 7.7 & 25.5 \\
 & inf 7B~\cite{infly-ai_2025} & 50.0 & 31.6 & 33.7 & 11.2 & 39.3 & 37.0 & 26.8 & 24.3 & 10.5 & 29.4 \\
 & sfr 7B~\cite{infly-ai_2025}  & 43.0 & 25.6 & 28.5 & 10.8 & 35.7 & 32.4 & 25.6 & 26.3 & 7.6 & 26.2 \\
\midrule
\multirow{7}{*}{InternVL3-2B} & splade 108M~\cite{lassance2024splade} & 32.1 & 23.8 & 12.0 & 14.8 & 33.8 & 22.9 & 8.4 & 10.6 & 5.8 & 18.3 \\
 & open search v1 133M~\cite{geng2024towards} & 29.7 & 35.6 & 15.3 & 13.4 & 33.5 & 21.9 & 7.3 & 10.3 & 5.0 & 19.1 \\
 & me5\_base 278M~\cite{wang2024multilingual} & 28.2 & 29.4 & 20.2 & 5.5 & 27.5 & 25.5 & 17.4 & 14.4 & 4.4 & 19.2 \\
 & me5\_large 560M~\cite{wang2024multilingual} & 39.9 & 19.2 & 20.5 & 5.9 & 31.3 & 30.3 & 26.5 & 24.3 & 5.1 & 22.6 \\
 & inf\_small 1.5B~\cite{infly-ai_2025} & 39.1 & 30.2 & 26.8 & 9.5 & 33.9 & 30.7 & 16.2 & 19.1 & 8.1 & 23.7 \\
 & inf 7B~\cite{infly-ai_2025} & 46.6 & 42.7 & 33.3 & 11.1 & 37.3 & 35.6 & 22.6 & 22.0 & 11.2 & 29.2 \\
 & sfr 7B~\cite{infly-ai_2025}  & 45.3 & 23.9 & 20.7 & 5.6 & 35.0 & 33.7 & 24.6 & 24.1 & 4.8 & 24.2 \\
\midrule
\multirow{7}{*}{InternVL3-8B} & splade 108M~\cite{lassance2024splade} & 37.9 & 8.2 & 9.9 & 9.2 & 37.2 & 23.7 & 14.6 & 15.4 & 3.7 & 17.8 \\
 & open search v1 133M~\cite{geng2024towards} & 35.8 & 9.6 & 12.7 & 7.3 & 34.2 & 22.9 & 17.3 & 15.6 & 3.1 & 17.6 \\
 & me5\_base 278M~\cite{wang2024multilingual} & 36.4 & 29.2 & 22.4 & 6.2 & 32.5 & 29.4 & 18.9 & 19.7 & 6.6 & 22.4 \\
 & me5\_large 560M~\cite{wang2024multilingual} & 32.5 & 9.3 & 22.1 & 6.5 & 33.0 & 32.4 & 23.7 & 26.0 & 6.0 & 21.3 \\
 & inf\_small 1.5B~\cite{infly-ai_2025} & 45.0 & 24.6 & 25.6 & 12.2 & 36.4 & 30.8 & 21.2 & 22.6 & 10.2 & 25.4 \\
 & inf 7B~\cite{infly-ai_2025} & 47.4 & 31.8 & 31.4 & 9.9 & 37.7 & 37.3 & 24.9 & 25.2 & 10.0 & 28.4 \\
 & sfr 7B~\cite{infly-ai_2025}  & 37.2 & 22.4 & 29.9 & 7.0 & 33.1 & 33.1 & 22.6 & 22.8 & 6.2 & 23.8 \\
\midrule
\multirow{7}{*}{InternVL3-14B} & splade 108M~\cite{lassance2024splade} & 43.5 & 11.2 & 9.8 & 6.6 & 36.5 & 23.6 & 17.0 & 15.6 & 2.6 & 18.5 \\
 & open search v1 133M~\cite{geng2024towards} & 37.9 & 11.7 & 11.1 & 7.2 & 36.3 & 23.2 & 12.1 & 13.2 & 2.7 & 17.3 \\
 & me5\_base 278M~\cite{wang2024multilingual} & 23.5 & 38.1 & 27.6 & 6.4 & 28.4 & 24.8 & 19.1 & 17.8 & 5.4 & 21.2 \\
 & me5\_large 560M~\cite{wang2024multilingual} & 42.0 & 21.4 & 21.2 & 8.3 & 31.4 & 29.9 & 25.8 & 26.0 & 7.3 & 23.7 \\
 & inf\_small 1.5B~\cite{infly-ai_2025} & 38.4 & 29.7 & 26.8 & 10.1 & 36.0 & 31.6 & 20.7 & 21.1 & 8.2 & 24.7 \\
 & inf 7B~\cite{infly-ai_2025} & 42.6 & 36.0 & 34.6 & 6.9 & 38.4 & 36.9 & 21.0 & 22.5 & 8.5 & 27.5 \\
 & sfr 7B~\cite{infly-ai_2025}  & 45.1 & 31.8 & 26.2 & 4.8 & 32.2 & 32.0 & 26.0 & 27.6 & 6.1 & 25.8 \\
 
\bottomrule
\bottomrule
\end{tabular}
\caption{Zeroshot VDR performance on ViDoRe-v2 benchmark. Metric: Recall@1.}
  \label{tab:results_r1}
  
\end{table*}

\begin{table*}
  \centering
  \tiny  
  \setlength{\tabcolsep}{3pt}
  \begin{tabular}{l l c c c c c c c c c | c}
\toprule
\toprule
\textbf{VLM} & \textbf{Text Encoder} & \textbf{RERB} & \textbf{SAXA} & \textbf{SAXAM} & \textbf{SEME} & \textbf{SMBTI} & \textbf{SMBTIM} & \textbf{SRS} & \textbf{SRSM} & \textbf{SEMEM} & \textbf{AVG} \\
\midrule
\multirow{7}{*}{Qwen2.5VL-7B} & splade 108M~\cite{lassance2024splade} & 69.2 & 26.3 & 27.1 & 30.2 & 61.4 & 43.0 & 35.6 & 38.2 & 13.4 & 38.3 \\
 & open search v1 133M~\cite{geng2024towards} & 63.2 & 30.5 & 29.2 & 31.3 & 63.5 & 45.6 & 38.4 & 38.4 & 15.6 & 39.5 \\
 & me5\_base 278M~\cite{wang2024multilingual} & 52.2 & 49.9 & 50.9 & 25.6 & 57.3 & 51.8 & 44.9 & 46.6 & 20.8 & 44.4 \\
 & me5\_large 560M~\cite{wang2024multilingual} & 69.3 & 58.8 & 57.3 & 27.0 & 62.1 & 59.6 & 56.4 & 55.6 & 24.5 & 52.3 \\
 & inf\_small 1.5B~\cite{infly-ai_2025} & 60.8 & 62.0 & 57.3 & 32.7 & 61.6 & 59.9 & 57.9 & 57.5 & 27.1 & 53.0 \\
 & inf 7B~\cite{infly-ai_2025} & 69.2 & 64.8 & 56.1 & 31.0 & 67.1 & 64.2 & 58.5 & 57.9 & 29.2 & 55.3 \\
 & sfr 7B~\cite{infly-ai_2025}  & 68.7 & 63.7 & 58.7 & 28.6 & 63.0 & 61.9 & 56.5 & 59.9 & 24.5 & 54.0 \\
\midrule
\multirow{7}{*}{Qwen2.5VL-32B} & splade 108M~\cite{lassance2024splade} & 74.9 & 48.1 & 36.0 & 30.3 & 65.6 & 44.2 & 39.8 & 37.7 & 15.1 & 43.5 \\
 & open search v1 133M~\cite{geng2024towards} & 75.9 & 51.9 & 38.5 & 30.5 & 65.5 & 45.2 & 37.9 & 36.0 & 14.8 & 44.0 \\
 & me5\_base 278M~\cite{wang2024multilingual} & 63.9 & 48.2 & 50.4 & 27.0 & 59.5 & 53.1 & 53.9 & 55.0 & 23.9 & 48.3 \\
 & me5\_large 560M~\cite{wang2024multilingual} & 73.1 & 56.1 & 52.5 & 25.6 & 63.1 & 60.9 & 53.0 & 55.0 & 23.4 & 51.4 \\
 & inf\_small 1.5B~\cite{infly-ai_2025} & 73.5 & 60.7 & 58.2 & 31.4 & 62.6 & 61.0 & 57.4 & 58.1 & 27.0 & 54.4 \\
 & inf 7B~\cite{infly-ai_2025} & 70.9 & 63.5 & 60.6 & 33.4 & 67.1 & 64.6 & 60.7 & 61.2 & 30.6 & 56.9 \\
 & sfr 7B~\cite{infly-ai_2025}  & 69.0 & 66.1 & 63.7 & 28.5 & 62.6 & 61.0 & 58.8 & 61.5 & 24.3 & 55.0 \\
\midrule
\multirow{7}{*}{Qwen2.5VL-72B} & splade 108M~\cite{lassance2024splade} & 73.0 & 11.1 & 23.1 & 31.4 & 64.1 & 45.3 & 36.1 & 39.4 & 15.7 & 37.7 \\
 & open search v1 133M~\cite{geng2024towards} & 68.3 & 17.6 & 27.1 & 32.2 & 63.2 & 46.1 & 36.9 & 39.3 & 16.6 & 38.6 \\
 & me5\_base 278M~\cite{wang2024multilingual} & 55.6 & 54.3 & 53.4 & 27.6 & 59.9 & 51.3 & 43.2 & 47.9 & 24.0 & 46.4 \\
 & me5\_large 560M~\cite{wang2024multilingual} & 70.6 & 50.9 & 53.8 & 27.3 & 64.6 & 62.3 & 48.6 & 49.0 & 26.5 & 50.4 \\
 & inf\_small 1.5B~\cite{infly-ai_2025} & 73.1 & 60.1 & 57.1 & 32.1 & 64.2 & 62.8 & 55.4 & 56.3 & 26.9 & 54.2 \\
 & inf 7B~\cite{infly-ai_2025} & 70.5 & 64.6 & 60.8 & 33.0 & 67.6 & 65.7 & 58.1 & 60.9 & 29.9 & 56.8 \\
 & sfr 7B~\cite{infly-ai_2025}  & 70.0 & 65.7 & 63.1 & 29.1 & 63.9 & 64.3 & 58.8 & 58.9 & 24.1 & 55.3 \\
\midrule
\multirow{7}{*}{InternVL3-2B} & splade 108M~\cite{lassance2024splade} & 58.0 & 50.8 & 28.8 & 29.2 & 61.2 & 42.7 & 31.9 & 37.9 & 15.7 & 39.6 \\
 & open search v1 133M~\cite{geng2024towards} & 54.5 & 56.9 & 32.1 & 31.3 & 62.4 & 42.9 & 34.9 & 34.8 & 17.0 & 40.8 \\
 & me5\_base 278M~\cite{wang2024multilingual} & 51.4 & 48.1 & 33.6 & 24.7 & 54.9 & 51.0 & 41.1 & 41.5 & 21.3 & 40.8 \\
 & me5\_large 560M~\cite{wang2024multilingual} & 65.5 & 55.8 & 40.9 & 26.4 & 61.1 & 59.1 & 50.3 & 51.0 & 23.1 & 48.1 \\
 & inf\_small 1.5B~\cite{infly-ai_2025} & 63.1 & 58.9 & 48.8 & 28.1 & 61.8 & 58.4 & 53.5 & 53.5 & 25.0 & 50.1 \\
 & inf 7B~\cite{infly-ai_2025} & 64.2 & 63.6 & 57.0 & 29.6 & 66.3 & 63.3 & 55.5 & 55.9 & 26.7 & 53.6 \\
 & sfr 7B~\cite{infly-ai_2025}  & 68.0 & 64.3 & 55.4 & 23.1 & 64.7 & 62.4 & 55.1 & 55.5 & 21.2 & 52.2 \\
\midrule
\multirow{7}{*}{InternVL3-8B} & splade 108M~\cite{lassance2024splade} & 76.5 & 30.1 & 30.4 & 30.0 & 61.7 & 43.5 & 39.3 & 41.8 & 15.6 & 41.0 \\
 & open search v1 133M~\cite{geng2024towards} & 68.3 & 28.2 & 32.6 & 31.6 & 62.5 & 44.8 & 39.5 & 41.9 & 14.9 & 40.5 \\
 & me5\_base 278M~\cite{wang2024multilingual} & 57.5 & 45.1 & 51.8 & 27.4 & 54.8 & 51.6 & 42.8 & 48.3 & 23.5 & 44.8 \\
 & me5\_large 560M~\cite{wang2024multilingual} & 66.3 & 43.4 & 49.4 & 27.0 & 63.5 & 60.7 & 58.7 & 59.8 & 23.9 & 50.3 \\
 & inf\_small 1.5B~\cite{infly-ai_2025} & 65.0 & 47.9 & 52.8 & 30.4 & 66.6 & 62.7 & 53.9 & 54.9 & 27.1 & 51.3 \\
 & inf 7B~\cite{infly-ai_2025} & 69.9 & 56.1 & 58.7 & 27.7 & 67.8 & 64.8 & 60.3 & 59.7 & 27.1 & 54.7 \\
 & sfr 7B~\cite{infly-ai_2025}  & 69.6 & 61.4 & 57.9 & 24.8 & 66.5 & 62.4 & 61.7 & 60.0 & 22.4 & 54.1 \\
\midrule
\multirow{7}{*}{InternVL3-14B} & splade 108M~\cite{lassance2024splade} & 75.3 & 40.4 & 32.4 & 29.3 & 62.3 & 43.7 & 39.1 & 38.6 & 14.1 & 41.7 \\
 & open search v1 133M~\cite{geng2024towards} & 69.6 & 39.9 & 32.7 & 30.1 & 61.8 & 44.6 & 36.5 & 37.4 & 14.1 & 40.7 \\
 & me5\_base 278M~\cite{wang2024multilingual} & 57.9 & 53.7 & 54.6 & 27.6 & 55.9 & 51.0 & 47.7 & 48.1 & 23.2 & 46.6 \\
 & me5\_large 560M~\cite{wang2024multilingual} & 72.8 & 58.4 & 55.5 & 26.6 & 61.8 & 59.2 & 55.4 & 58.1 & 22.4 & 52.2 \\
 & inf\_small 1.5B~\cite{infly-ai_2025} & 66.1 & 59.3 & 56.0 & 29.8 & 64.1 & 61.4 & 56.3 & 56.5 & 26.5 & 52.9 \\
 & inf 7B~\cite{infly-ai_2025} & 73.6 & 58.8 & 55.8 & 32.9 & 66.8 & 64.9 & 54.9 & 56.7 & 29.7 & 54.9 \\
 & sfr 7B~\cite{infly-ai_2025}  & 69.2 & 64.1 & 60.0 & 28.6 & 64.6 & 63.5 & 54.3 & 55.6 & 23.7 & 53.7 \\
\bottomrule
\bottomrule
\end{tabular}
\caption{Zeroshot VDR performance on ViDoRe-v2 benchmark. Metric: Recall@5.}
  \label{tab:results_r5}
  
\end{table*}

\begin{table*}
  \centering
  \tiny 
  \setlength{\tabcolsep}{3pt}
  \begin{tabular}{l l c c c c c c c c c | c}
\toprule
\toprule
\textbf{VLM} & \textbf{Text Encoder} & \textbf{RERB} & \textbf{SAXA} & \textbf{SAXAM} & \textbf{SEME} & \textbf{SMBTI} & \textbf{SMBTIM} & \textbf{SRS} & \textbf{SRSM} & \textbf{SEMEM} & \textbf{AVG} \\
\midrule
\multirow{7}{*}{Qwen2.5VL-7B} & splade 108M~\cite{lassance2024splade} & 77.1 & 30.9 & 32.9 & 44.7 & 72.6 & 52.4 & 50.7 & 53.4 & 23.7 & 48.7 \\
 & open search v1 133M~\cite{geng2024towards} & 74.9 & 36.1 & 35.5 & 42.4 & 72.0 & 53.4 & 49.8 & 48.9 & 23.5 & 48.5 \\
 & me5\_base 278M~\cite{wang2024multilingual} & 64.5 & 58.2 & 57.2 & 39.7 & 66.9 & 62.4 & 63.4 & 63.0 & 32.4 & 56.4 \\
 & me5\_large 560M~\cite{wang2024multilingual} & 77.5 & 64.4 & 64.5 & 38.4 & 76.2 & 72.2 & 68.7 & 69.1 & 35.5 & 63.0 \\
 & inf\_small 1.5B~\cite{infly-ai_2025} & 75.3 & 67.8 & 66.3 & 44.2 & 75.9 & 72.0 & 70.1 & 72.0 & 39.5 & 64.8 \\
 & inf 7B~\cite{infly-ai_2025} & 74.9 & 71.1 & 67.0 & 45.0 & 77.1 & 74.8 & 71.5 & 69.9 & 42.3 & 65.9 \\
 & sfr 7B~\cite{infly-ai_2025}  & 76.3 & 72.6 & 72.7 & 39.6 & 73.7 & 72.5 & 73.0 & 76.3 & 37.0 & 66.0 \\
\midrule
\multirow{7}{*}{Qwen2.5VL-32B} & splade 108M~\cite{lassance2024splade} & 88.1 & 59.7 & 49.9 & 44.6 & 74.8 & 54.0 & 55.7 & 53.6 & 24.2 & 56.1 \\
 & open search v1 133M~\cite{geng2024towards} & 86.2 & 66.8 & 53.0 & 45.5 & 73.8 & 55.2 & 56.5 & 52.0 & 25.3 & 57.1 \\
 & me5\_base 278M~\cite{wang2024multilingual} & 71.0 & 57.3 & 60.1 & 37.6 & 71.0 & 65.0 & 69.5 & 67.0 & 33.5 & 59.1 \\
 & me5\_large 560M~\cite{wang2024multilingual} & 80.4 & 62.9 & 59.9 & 37.2 & 73.8 & 71.6 & 67.1 & 67.3 & 35.5 & 61.7 \\
 & inf\_small 1.5B~\cite{infly-ai_2025} & 78.2 & 70.6 & 66.4 & 45.5 & 74.3 & 71.3 & 70.4 & 73.4 & 39.5 & 65.5 \\
 & inf 7B~\cite{infly-ai_2025} & 79.9 & 71.6 & 69.0 & 44.2 & 76.7 & 75.2 & 77.4 & 77.4 & 42.0 & 68.2 \\
 & sfr 7B~\cite{infly-ai_2025}  & 76.3 & 71.5 & 72.6 & 38.8 & 74.0 & 72.4 & 74.0 & 76.4 & 35.6 & 65.7 \\
\midrule
\multirow{7}{*}{Qwen2.5VL-72B} & splade 108M~\cite{lassance2024splade} & 83.8 & 23.1 & 34.6 & 42.8 & 74.5 & 54.6 & 51.4 & 54.5 & 25.8 & 49.5 \\
 & open search v1 133M~\cite{geng2024towards} & 80.3 & 32.8 & 38.8 & 45.0 & 73.8 & 54.6 & 55.4 & 53.3 & 25.4 & 51.0 \\
 & me5\_base 278M~\cite{wang2024multilingual} & 64.2 & 61.1 & 59.9 & 40.4 & 70.8 & 63.6 & 62.6 & 63.1 & 34.7 & 57.8 \\
 & me5\_large 560M~\cite{wang2024multilingual} & 77.3 & 63.8 & 63.2 & 38.6 & 74.5 & 73.0 & 65.0 & 65.4 & 36.6 & 61.9 \\
 & inf\_small 1.5B~\cite{infly-ai_2025} & 78.7 & 69.5 & 65.2 & 44.8 & 74.3 & 72.8 & 70.7 & 73.3 & 40.0 & 65.5 \\
 & inf 7B~\cite{infly-ai_2025} & 76.3 & 74.4 & 67.6 & 45.4 & 77.4 & 76.7 & 76.7 & 78.0 & 43.8 & 68.5 \\
 & sfr 7B~\cite{infly-ai_2025}  & 82.7 & 74.3 & 71.4 & 40.4 & 78.2 & 74.7 & 71.5 & 75.8 & 37.0 & 67.3 \\
\midrule
\multirow{7}{*}{InternVL3-2B} & splade 108M~\cite{lassance2024splade} & 73.1 & 62.5 & 35.0 & 44.2 & 73.0 & 52.7 & 51.4 & 56.1 & 24.6 & 52.5 \\
 & open search v1 133M~\cite{geng2024towards} & 67.9 & 65.0 & 37.7 & 43.7 & 72.0 & 53.0 & 51.1 & 51.3 & 25.2 & 51.9 \\
 & me5\_base 278M~\cite{wang2024multilingual} & 61.6 & 60.0 & 44.2 & 37.1 & 66.9 & 61.4 & 50.8 & 52.2 & 32.8 & 51.9 \\
 & me5\_large 560M~\cite{wang2024multilingual} & 76.0 & 67.2 & 54.3 & 37.7 & 71.9 & 69.2 & 58.4 & 61.5 & 34.6 & 59.0 \\
 & inf\_small 1.5B~\cite{infly-ai_2025} & 68.8 & 64.7 & 58.3 & 42.9 & 73.6 & 69.5 & 67.9 & 69.1 & 38.8 & 61.5 \\
 & inf 7B~\cite{infly-ai_2025} & 76.7 & 72.9 & 67.5 & 43.2 & 77.3 & 74.7 & 68.5 & 68.6 & 39.9 & 65.5 \\
 & sfr 7B~\cite{infly-ai_2025}  & 73.9 & 72.0 & 63.1 & 34.9 & 76.1 & 73.2 & 69.7 & 71.8 & 32.6 & 63.0 \\
\midrule
\multirow{7}{*}{InternVL3-8B} & splade 108M~\cite{lassance2024splade} & 81.0 & 36.9 & 41.6 & 44.7 & 72.3 & 53.3 & 52.9 & 57.0 & 26.1 & 51.7 \\
 & open search v1 133M~\cite{geng2024towards} & 80.4 & 48.7 & 44.0 & 42.9 & 73.3 & 54.2 & 55.3 & 57.8 & 24.3 & 53.5 \\
 & me5\_base 278M~\cite{wang2024multilingual} & 71.6 & 50.8 & 58.4 & 39.5 & 66.9 & 61.9 & 63.7 & 63.6 & 33.5 & 56.6 \\
 & me5\_large 560M~\cite{wang2024multilingual} & 79.0 & 52.4 & 61.1 & 37.1 & 74.2 & 71.0 & 74.0 & 73.2 & 34.5 & 61.8 \\
 & inf\_small 1.5B~\cite{infly-ai_2025} & 72.8 & 58.8 & 63.0 & 43.8 & 75.5 & 71.7 & 69.9 & 70.9 & 39.8 & 62.9 \\
 & inf 7B~\cite{infly-ai_2025} & 81.3 & 71.2 & 67.4 & 44.4 & 77.5 & 75.7 & 74.5 & 74.5 & 41.4 & 67.5 \\
 & sfr 7B~\cite{infly-ai_2025}  & 82.4 & 71.5 & 68.0 & 36.0 & 75.3 & 74.1 & 74.4 & 76.7 & 34.2 & 65.8 \\
\midrule
\multirow{7}{*}{InternVL3-14B} & splade 108M~\cite{lassance2024splade} & 80.9 & 56.3 & 44.4 & 45.2 & 72.7 & 53.2 & 50.9 & 53.8 & 24.9 & 53.6 \\
 & open search v1 133M~\cite{geng2024towards} & 77.5 & 62.9 & 47.0 & 45.3 & 72.3 & 54.1 & 50.5 & 53.5 & 24.9 & 54.2 \\
 & me5\_base 278M~\cite{wang2024multilingual} & 72.8 & 60.8 & 65.8 & 36.8 & 66.1 & 61.6 & 60.0 & 60.8 & 33.1 & 57.5 \\
 & me5\_large 560M~\cite{wang2024multilingual} & 82.7 & 66.7 & 65.2 & 38.7 & 72.5 & 69.9 & 68.8 & 70.6 & 34.1 & 63.2 \\
 & inf\_small 1.5B~\cite{infly-ai_2025} & 74.0 & 68.7 & 67.4 & 45.1 & 75.1 & 72.6 & 71.8 & 71.0 & 40.2 & 65.1 \\
 & inf 7B~\cite{infly-ai_2025} & 80.1 & 68.9 & 69.1 & 45.0 & 76.9 & 75.6 & 72.7 & 72.4 & 42.7 & 67.0 \\
 & sfr 7B~\cite{infly-ai_2025}  & 77.0 & 69.7 & 70.2 & 39.8 & 75.3 & 73.8 & 71.0 & 73.3 & 36.7 & 65.2 \\
 
\bottomrule
\bottomrule
\end{tabular}
\caption{Zeroshot VDR performance on ViDoRe-v2 benchmark. Metric: Recall@10.}
  \label{tab:results_r10}
  
\end{table*}


\end{document}